\documentclass[twocolumn,twocolappendix]{aastex63}

\graphicspath{{../code/figs/}}
\usepackage{comment}
\usepackage{amsmath,amssymb}
\usepackage{bm}

\newcommand{\msun}{M_\odot}
\newcommand{\rsun}{R_\odot}

\renewcommand{\d}{\mathrm{d}}

\newcommand{\g}{\mathcal{N}}
\renewcommand{\u}{\mathcal{U}}
\newcommand{\logu}{\mathcal{U}_\mathrm{log}}

\shorttitle{Geometry of the KOI-89 System}
\shortauthors{Masuda and Tamayo}

\begin{document}


\title{
Revisiting the Architecture of the KOI-89 System
}

\correspondingauthor{Kento Masuda}
\email{kmasuda@ess.sci.osaka-u.ac.jp}

\author[0000-0003-1298-9699]{Kento Masuda}
\affiliation{Department of Earth and Space Science, Osaka University, Osaka 560-0043, Japan}

\author[0000-0002-9908-8705]{Daniel Tamayo}
\altaffiliation{NASA Sagan Fellow}
\affiliation{Department of Astrophysical Sciences, Princeton University, Princeton, NJ 08544, USA}

\received{July 28, 2020}
\revised{September 11, 2020}
\accepted{September 14, 2020}




\begin{abstract}
While high stellar obliquities observed in exoplanetary systems may be attributed to processes that tilt the planetary orbits, it is also possible that they reflect misalignments between protoplanetary disks and stellar spins. This latter hypothesis predicts the presence of co-planar multi-planetary systems misaligned with their central stars.
Here we re-evaluate the evidence of such an architecture that has been claimed for the KOI-89 system. KOI-89 is an early-type star with one validated transiting planet KOI-89.01/Kepler-462b (period 84.7~days, radius $3.0\,R_\oplus$) and one transiting planet candidate KOI-89.02 (period 207.6~days, radius $4.0\,R_\oplus$), where the latter exhibits transit timing variations (TTVs). A previous modeling of the stellar gravity-darkening effect in the transit light curves inferred a high stellar obliquity of $\approx70^\circ$. We perform a photodynamical modeling of the \textit{Kepler} transit light curves, and use the resulting constraints on the orbital configuration and transit times to update the gravity-darkened transit model. As a result, we find no firm evidence for gravity darkening effect in the transit shapes and conclude that stellar obliquity is not constrained by the data. Given evidence for low orbital eccentricities from the dynamical analysis, the system architecture can thus be consistent with many other multi-transiting systems with flat, near-circular orbits aligned with the stellar spin. We find that the TTVs imparted on its neighbor imply that KOI-89.01 has a mass $\gtrsim20\,M_\oplus$. This would render it one of the densest known sub-Neptunes, mostly composed of a solid core. Lower masses are possible if the TTVs are instead due to an unseen third planet.
\end{abstract}



\section{Introduction}\label{sec:intro}

The approximate alignment between the Sun's rotation and the orbital motion of our Solar System's planets appears to be a natural consequence of planet formation in a protoplanetary disk. Yet, measurements of stellar obliquities in transiting explanetary systems have revealed that this is not always the case. Stellar obliquities have been constrained in more than 100 systems\footnote{A summary can be found in TEPCAT \citep[\url{https://www.astro.keele.ac.uk/jkt/tepcat/tepcat.html;}][]{2011MNRAS.417.2166S}.}, and large stellar obliquities have been reported not only for hot Jupiters \citep[e.g.,][]{2008A&A...488..763H} but also for 
planets smaller than about Neptune \citep[e.g.,][]{2010ApJ...723L.223W, 2018Natur.553..477B,2019AJ....157..137K,2020arXiv200711564K}

It remains an open question
whether the planets are to blame or not for those misalignments.
They may originate from post-formation perturbations due to planet-planet scattering \citep{1996Sci...274..954R, 1996Natur.384..619W, 2008ApJ...686..580C, 2008ApJ...686..603J} or 
secular interactions with a stellar/planetary companion
\citep{1997ApJ...477L.103M, 1997Natur.386..254H, 2007ApJ...669.1298F, 2011Natur.473..187N}.
If this is the case, stellar obliquities would serve as a probe of dynamical evolution of planetary orbits.

Alternatively, high stellar obliquities may simply reflect primordial misalignments between stellar equators and their surrounding protoplanetary disks:
disks may be born misaligned with their host stars due to chaotic accretion of angular momentum expected from a turbulent molecular cloud \citep{2010MNRAS.401.1505B, 2015MNRAS.450.3306F, 2020MNRAS.492.5641T},
or disks may be tilted after the formation due to (combined effects of) torques from an exterior stellar companion and gravitational/magnetic star--disk interactions \citep{2012Natur.491..418B, 2013ApJ...778..169B, 2014ApJ...790...42S, 2014MNRAS.440.3532L, 2015ApJ...811...82S, 2018MNRAS.478..835Z}.
Alternatively, a massive star's surface rotation might be reoriented after disk/planet formation due to angular momentum transport via internal gravity waves \citep{2012ApJ...758L...6R}.

To test this latter possibility, two complementary methods have been pursued. First, one can search for misalignments between stars and their protoplanetary disks. Because stars fully embedded in their protoplanetary disks cannot be observed, such measurements have been performed for resolved debris disks as proxies of the original protoplanetary disks \citep{2011MNRAS.413L..71W, 2014MNRAS.438L..31G}, and for protoplanetary disks in the late phase of evolution \citep{2019MNRAS.484.1926D}. While the majority of theses disks appear to be star-aligned \citep{2011MNRAS.413L..71W, 2014MNRAS.438L..31G}, some candidates for misaligned systems have also been presented \citep{2019MNRAS.484.1926D}. A possibly confounding factor is that the inner disk regions more relevant to the observed exoplanet population could be aligned, even though the resolved outer regions are misaligned.

The second possibility, which we focus on in this paper, is to use nearly coplanar multi-planet systems, whose orbits appear to have remained intact after formation. Obliquities have been measured for a dozen of stars with multiple transiting planets to date \citep{2012Natur.487..449S,2012ApJ...759L..36H,2013ApJ...771...11A,2013ApJ...766..101C,2013Sci...342..331H,2015ApJ...812L..11S,2015ApJ...814...67A,2018AJ....155...70W,2018AJ....156...93Z,2019A&A...631A..28D,2020ApJ...890L..27H,2020arXiv200500047M}. These multi-transiting systems most likely \citep[though not always;][]{2017AJ....153...45M} have aligned orbital planes that trace the initial protoplanetary disk.

While most of these measurements are consistent with low ($\lesssim10^\circ$) stellar obliquity,
three of these systems, Kepler-56 \citep{2013Sci...342..331H}, KOI-89 \citep{2015ApJ...814...67A}, and HD~3167 \citep{2019A&A...631A..28D} have strong claimed misalignments. 
If the planets in these systems trace their natal disk, they are smoking guns for 
an obliquity excitation process independent of orbital evolution.

However, the Kepler-56 planets may not trace the original plane of the protoplanetary disk. There is in fact a third massive ($\gtrsim5.5\,M_{\rm Jup}$) planet on an au-scale, eccentric orbit \citep{2016AJ....152..165O} exterior to the two transiting planets. Follow-up studies have shown that gravitational interactions with the outer planet could have tilted the inner orbits out of alignment with the host star while preserving their mutual coplanarity, owing to the tight gravitational coupling between the inner transiting planets \citep{2014ApJ...794..131L,2017AJ....153...42L,2017MNRAS.464.1709G, 2017AJ....153..210H}. \citet{2019A&A...631A..28D} argued that the same could well be the case for HD~3167 too: they reported a possible signature of an outer companion in the radial velocity data, and showed that the hypothetical companion is physically capable of generating a high stellar obliquity while maintaining the co-planarity of the inner system.\footnote{Although these explanations require that the outer giant has a misaligned orbit relative to the inner planets, there has been accumulating observational evidence that such misaligned systems exist \citep{2020AJ....159...38M, 2020MNRAS.tmp.2175X, 2020arXiv200708549D, 2020arXiv200706410D}.}

That leaves KOI-89 as the best candidate for misalignments unrelated to orbital evolution, including primordial star--disk misalignment.
Even if undetected companions existed in the system, the transiting planets are more widely separated than in the above systems
and are thus less likely to maintain co-planarity under plausible external perturbations. 
In this light, it is particularly important to evaluate the evidence for the claimed high obliquity of KOI-89, and/or to search for potential signature of mutual orbital misalignment.

The high ($\approx 70^\circ$) stellar obliquity of KOI-89 was inferred \citep{2015ApJ...814...67A} by modeling the gravity-darkening effect \citep{1924MNRAS..84..665V} in transit light curves \citep{2009ApJ...705..683B}. However, the solution involved a large ($\gtrsim 0.5$) orbital eccentricity for KOI-89.02 that may make its orbit cross with the inner one, suggesting that there is something missing in the modeling. A wrong inference for the eccentricity could also bias the transit impact parameters and hence the obliquity measurement. In addition, KOI-89.02 exhibits strong transit timing variations (TTVs) that can further complicate the analysis of the transit shape. 

In this paper, we revisit the geometric architecture of the KOI-89 system, in particular the constraint on stellar obliquity, by fully exploiting the information from transit shapes and TTVs.
In Section \ref{sec:star}, we characterize the host star combining archival photometry, spectroscopy, and astrometry data. 
Given the large number of additional parameters required to model transits across a potentially tilted, gravity-darkened star, we begin with a photodynamical analysis ignoring the effect of rapid stellar rotation (Section \ref{sec:photod}).
We then fold in the resulting constraints on the orbital configuration to evaluate the evidence for a non-zero stellar obliquity in Section \ref{sec:gravity_darkening}.
From these analyses, we find that the stellar obliquity is not well constrained and that the architecture of the system can in fact be consistent with a flat, near-circular geometry as inferred for most other multi-transiting systems. 
Section \ref{sec:discussion} puts this updated measurement into context, and discusses the potentially interesting physical nature of the planets inferred from our photodynamical modeling.

\begin{figure*}
	\plotone{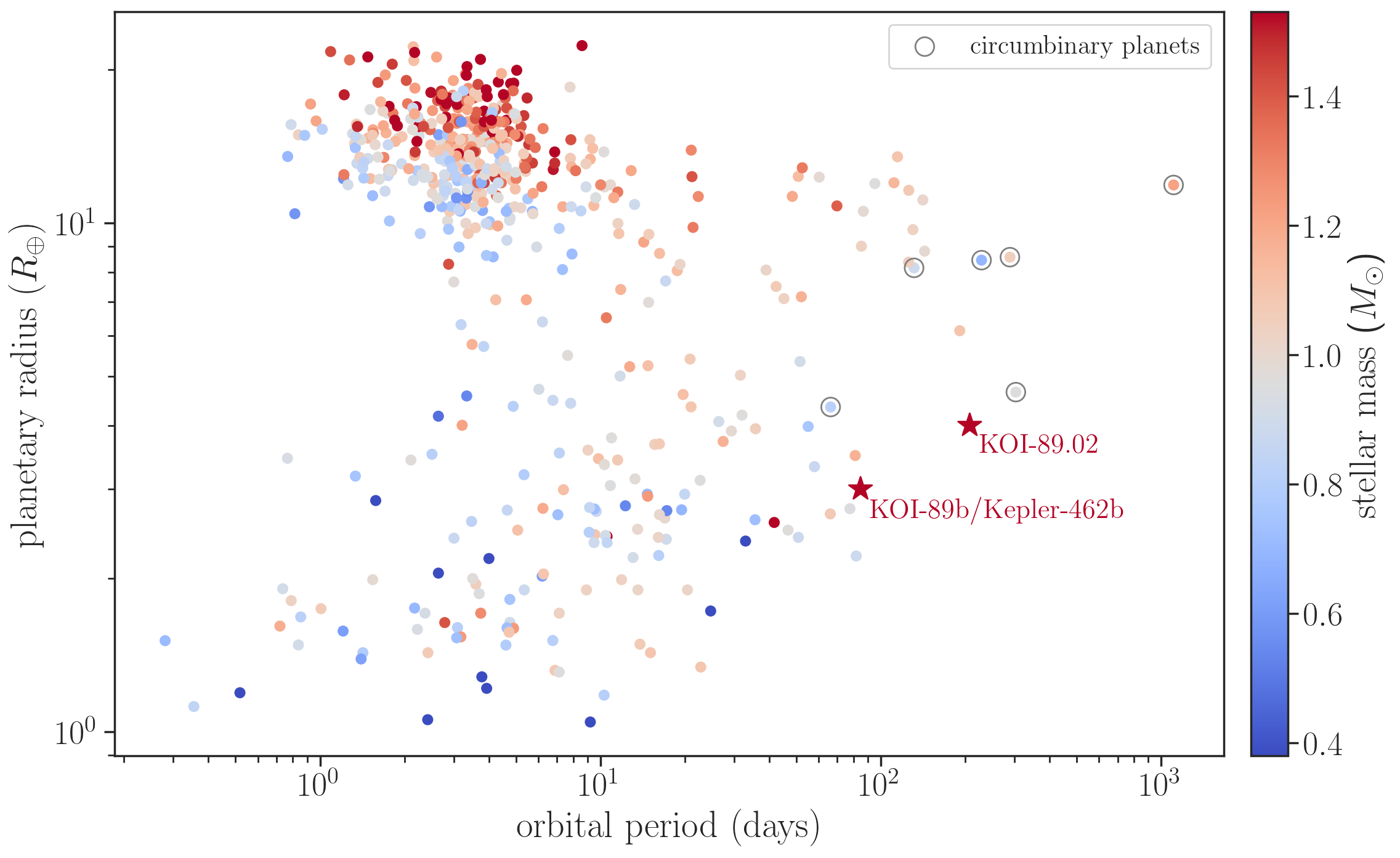}
	\caption{KOI-89 planets are among the longest-period, smallest transiting planets with mass constraints, especially around single host stars more massive than the sun. The colored circles are radii and orbital periods of planets with mass constraints from the NASA exoplanet archive, where planets without reported uncertainties and those from direct imaging observations are not shown. Planets orbiting two stars are marked with outer gray circles.}
	\label{fig:radius_period_smass}
\end{figure*}

\section{Stellar Parameters}\label{sec:star}

We characterized the host star KOI-89 by comparing stellar evolutionary models \citep{2016ApJS..222....8D, 2016ApJ...823..102C} computed with Modules for Experiments in Stellar Astrophysics \citep[MESA;][]{2011ApJS..192....3P,2013ApJS..208....4P, 2015ApJS..220...15P} to the atmospheric parameters from spectroscopy, the apparent $K$-band magnitude, and the distance from {\it Gaia} Data Release 2 \citep{2018A&A...616A...1G}. We used the {\tt isochrones} package by \citet{2015ascl.soft03010M} to obtain posterior samples for the stellar parameters. The same procedure was adopted in \citet{2018ApJ...864L..38D} and was validated using a sample of stars characterized with asteroseismology. The results are summarized in Table \ref{tab:star}. The host star KOI-89 is an early-type main-sequence star.

The atmospheric parameters were adopted from the analysis of Large Sky Area Multi-Object Fiber Spectroscopic Telescope \citep[LAMOST;][]{2012RAA....12.1197C, 2015RAA....15.1095L} spectra by \citet{2016A&A...594A..39F}: $T_{\rm eff}=7580\pm300\,\mathrm{K}$, $\log g=3.88\pm0.3$, $\mathrm{[Fe/H]}=0.13\pm0.2$, where the uncertainties are based on the externally estimated accuracies of the pipeline \citep{2016A&A...594A..39F}. The $K$-magnitudes were taken from the Two Micron All Sky Survey \citep[2MASS;][]{2006AJ....131.1163S}. We used the dust map and extinction vector from {\tt Bayestar17} \citep{2018MNRAS.478..651G} to correct for extinction in the $K$-band, assuming a 30\% fractional uncertainty \citep{2018AJ....156..264F}. The distance is based on \citet{2018AJ....156...58B} rather than the inverse of the parallax.

The derived mass and radius are consistent with those in the version 8 of the {\it TESS} Input Catalog \citep[TIC;][]{2019AJ....158..138S}: $M_\star=1.8\pm0.3\,\msun$ and $R_\star=1.62\pm0.06\,\rsun$. On the other hand, our radius (as well as that in TIC) is slightly smaller than $1.74^{+0.08}_{-0.06}\,\rsun$ by \citet{2018ApJ...866...99B}, essentially because they adopted a lower effective temperature than we did ($T_{\rm eff}=6688\pm134\,\mathrm{K}$,  $\log g=4.059\pm0.150$, $\mathrm{[Fe/H]}=-0.210\pm0.150$), which originates from the {\it Kepler} Input Catalog DR 25 \citep{2017ApJS..229...30M}. The value was originally derived in \citet{2014ApJ...784...45R} using the specmatch algorithm on the Keck HIRES spectrum \citep{2013ApJ...770...69P}. 
However, this lower temperature is not supported by the empirical color--temperature relation from \citet{2013ApJ...771...40B}, which gives $T_{\rm eff}\approx7000\,\mathrm{K}$ for $V-J$, $V-H$, $V-K$ even without correcting for extinction; the true color of the star should be bluer, and so the true temperature should be higher. 
We also reduced the archival Keck HIRES spectrum\footnote{Downloaded from the Keck observatory archive. \url{https://koa.ipac.caltech.edu/cgi-bin/KOA/nph-KOAlogin}} using the {\tt CERES} pipeline \citep{2017PASP..129c4002B} and derived stellar atmospheric parameters applying the {\tt ZASPE} \citep{2017MNRAS.467..971B} code, in which we extended the default temperature range using a synthetic spectrum library based on the PHOENIX model atmospheres \citep{2013A&A...553A...6H}. We found $T_{\rm eff}=7200\pm340\,\mathrm{K}$, $\mathrm{[Fe/H]}=-0.04\pm0.18$, and $v\sin i = 80\pm3\,\mathrm{km/s}$ when we imposed the constraint $\log g=4.14\pm0.1$ which is in between the values from \citet{2014ApJ...784...45R} and \citet{2016A&A...594A..39F}, and is also consistent with the TIC value. Thus we conclude that $T_{\rm eff}\approx7500\,\mathrm{K}$ we adopted is reasonable. The cause of the discrepancy in the \citet{2014ApJ...784...45R} measurement using the same Keck spectrum remains unclear, but this might be because specmatch was not designed for early-type, rapidly rotating stars.

\begin{deluxetable}{lccc}
\caption{Parameters of the Host Star KOI-89/Kepler-462.}
\label{tab:star}
\tablehead{
	\colhead{Parameter} & \colhead{Value}
} 
\startdata
effective temperature (K) & $7500\pm300$\\
log surface gravity (cgs)	& $4.22\pm0.04$\\
metallicity (dex)	& $0.05\pm0.11$\\
distance (pc)	& $598\pm11$\\
mass $M_\star$ $(M_\odot)$ & $1.59\pm0.08$\\
radius $R_\star$ $(R_\odot)$ & $1.62\pm0.04$\\
mean density $\rho_\star$ $(\mathrm{g\,cm^{-3}})$ & $0.53\pm0.06$\\
$\log_{10}$ age (yr)	& $8.7\pm0.4$\\
projected rotation velocity $v\sin i_\star$ $(\mathrm{km\,s^{-1}})$ & $80\pm3$\\
2MASS $K_s$-magnitude & $10.85\pm0.02$\\
\enddata
\end{deluxetable}

\section{Photodynamical Modeling}\label{sec:photod}

Here we model the {\it Kepler} transit light curves of KOI-89.01/Kepler-462b and KOI-89.02 taking into account their gravitational interactions, and constrain their masses, radii, orbital eccentricities, and orbital inclinations. The analysis establishes the planetary status of KOI-89.02, and shows that the orbits of the two planets are nearly circular and well-aligned. We also check how much the results can change if the TTVs of KOI-89.02 are caused by an outer, undetected planet. 
Because the analysis here essentially focuses on transit times and durations, we ignore the stellar oblateness and gravity darkening assuming that their effects on the relevant parameters are minor. We will check the consistency of this assumption later in Section \ref{sec:gravity_darkening}.

\subsection{The Data and Detrending}\label{ssec:photod_data}

We analyzed the long-cadence, Pre-search Data Conditioning (PDC) light curves downloaded from the Mikulski Archive for Space Telescopes.\footnote{\url{https://archive.stsci.edu}} The transits of each planet were fitted iteratively as described in \citet{2015ApJ...805...28M}, using the analytic light curve model for the quadratic limb-darkening law \citep{2002ApJ...580L.171M} as implemented in {\tt batman} \citep{2015PASP..127.1161K}, multiplied by a quadratic polynomial function of time that accounts for longer-term trends. We fit both transit times and mid-durations letting impact parameters to vary for each transit, and the fitted transit times were used to stack transits without binning; then the stacked transits were fitted to update transit-shape parameters for the next iteration. After 10 iterations, the light curve around each transit was divided by the best-fit polynomial. We use these normalized and detrended transit light curves for photodynamical modeling. An overlapping transit around $\mathrm{BJD}_\mathrm{TDB}=2454833+912.7$ was omitted from this analysis, but was later used for a consistency check.

\subsection{The Model}\label{ssec:photod_model}

We fit osculating Jacobi elements defined at $\mathrm{BJD}_{\mathrm{TBD}}=2454833+149$ and mass ratios for each planet, as well as quadratic limb-darkening coefficients \citep[as parameterized in][]{2013MNRAS.435.2152K} and mean density of the star.\footnote{Because the light-curve model alone does not constrain masses and lengths separately, we fix the stellar mass to be $1\,\msun$ and fit only the density of the star as well as ratios of masses and radii of the star and planets without loss of generality.}
We used {\tt TTVFast} \citep{2014ApJ...787..132D} to compute model mid-transit time, sky-plane velocity, and impact parameter during each transit of each planet. They were used to calculate planet locations on the stellar surface assuming a linear function of time,\footnote{Ignoring the change in the sky-projected velocity during transits has a negligible effect for such long-period planets.} and the locations, along with planet-to-star radius ratios and limb-darkening coefficients, were used to calculate the relative flux losses. We again adopted the \citet{2002ApJ...580L.171M} model for a quadratically limb-darkened star, as implemented in {\tt batman} \citep{2015PASP..127.1161K}, and computed the long-cadence fluxes $\bm{m}$ with a supersampling factor of 11. The residuals between the data $\bm{f}$ and the transit model $\bm{m}$ were modeled as a Gaussian process whose covariance matrix consists of a Mat\'ern-3/2 covariance and a white-noise term. Thus the log-likelihood of the model $\ln\mathcal{L}$ is
\begin{equation}
    \ln\mathcal{L} = -{1\over2}(\bm{f}-\bm{m})^T K^{-1} (\bm{f}-\bm{m})-{1\over 2}\ln\mathrm{det}\,K + \mathrm{const.},
\end{equation}
where 
\begin{equation}
    \label{eq:kernel}
    K_{ij}=(\sigma_i^2+\sigma_{\rm jit}^2)\delta_{ij}
    +\alpha^2 \left(1+{|t_i-t_j|\over 3\rho}\right)
    \exp\left(-{|t_i-t_j|\over 3\rho}\right).
\end{equation}
Here $t_i$ and $\sigma_i$ are the time and PDC flux error of the $i$th data point, respectively.
This likelihood was evaluated using {\tt celerite} \citep{celerite}.

We assumed a Gaussian prior for the mean stellar density with a central value of $0.53\,\mathrm{g\,cm^{-3}}$ and a width of $0.06\,\mathrm{g\,cm^{-3}}$ based on the results in Section \ref{sec:star}. For the other parameters, we adopted uniform or log-uniform priors on the parameters as specified in Table \ref{tab:koi89}. The posterior samples were obtained using the nested-sampling algorithm {\tt MultiNest} \citep{2008MNRAS.384..449F, 2009MNRAS.398.1601F, 2013arXiv1306.2144F} and its python interface {\tt PyMultiNest} \citep{2014A&A...564A.125B}. We typically used 4000 live points and a sampling efficiency of 0.8, and kept updating the live points until an evidence tolerance of 0.5 is achieved. We allowed for the detection of multiple posterior modes.

\subsection{Results: Two-planet Model}\label{ssec:photod_two}

As the simplest possible explanation, we first adopt a two-planet model.
We first fit the data assuming log-uniform priors for the planet-to-star mass ratios $m_1/M_\star$ and $m_2/M_\star$, and a truncated uniform prior $e_1<0.8$ and $e_2<0.45$ for the orbital eccentricities, where the subscripts 1 and 2 denote KOI-89.01 and KOI-89.02, respectively. These prior limits on $e_1$ and $e_2$ ensure that the orbits of the two planets do not cross, i.e., the periastron distance of the outer orbit is larger than the apastron distance of the inner orbit.
Crossing orbital configurations lead to rapid instabilities and are thus short-lived.

\begin{figure*}
\epsscale{1.2}
\plotone{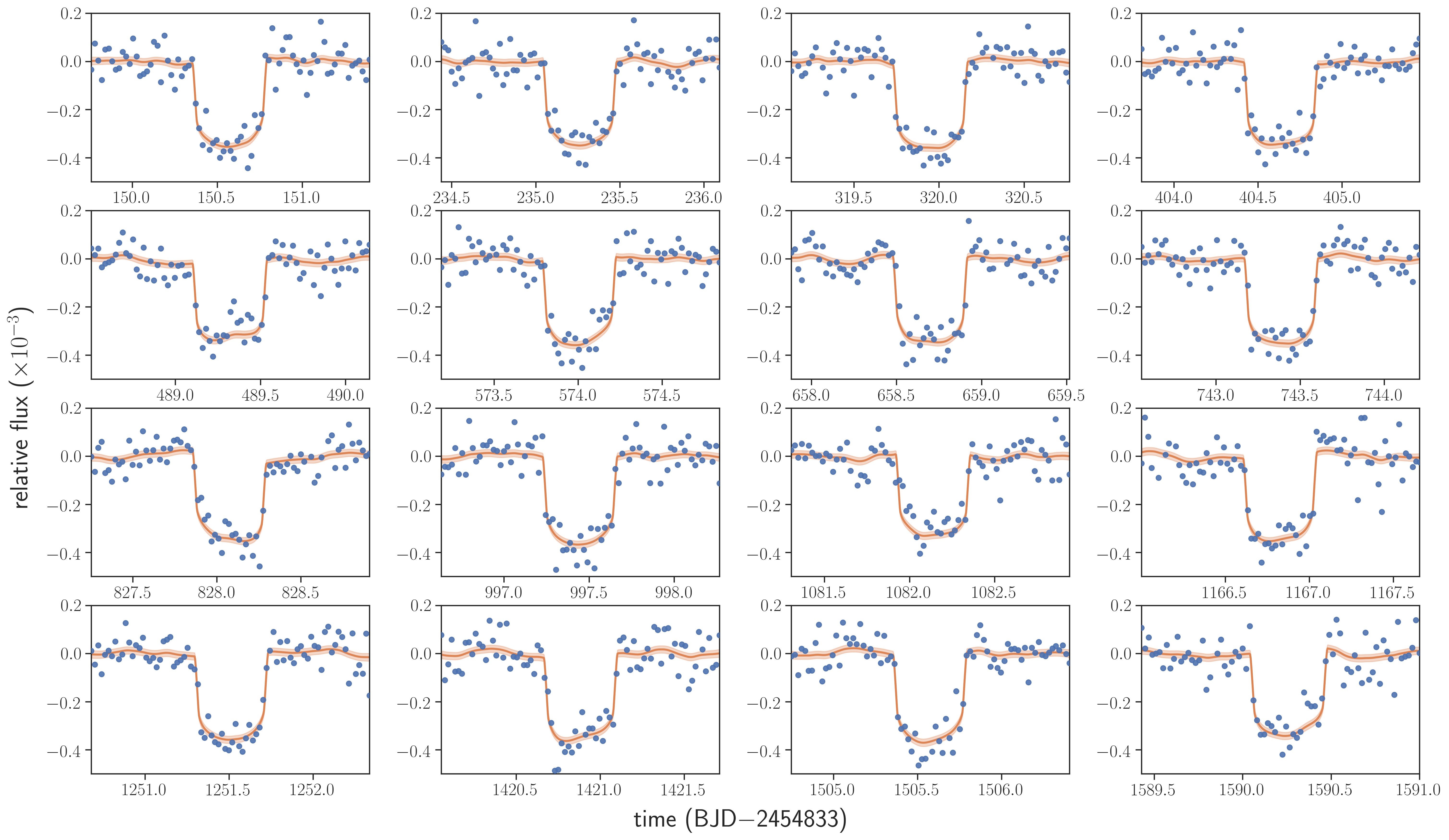}
\caption{Transit light curves and posterior models for KOI-89.01. The blue filled circles show the observed flux values, and the orange lines show the maximum likelihood model with its standard deviation (shaded region).}
\label{fig:transits_01}
\end{figure*}

\begin{figure*}
\epsscale{1.2}
\plotone{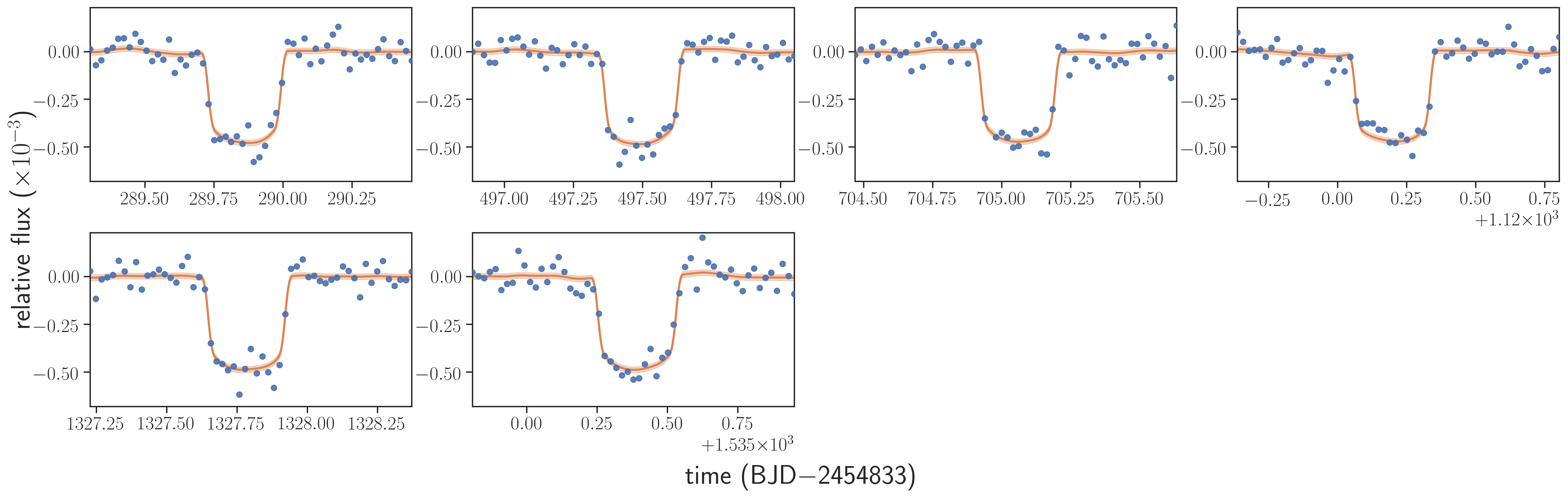}
\caption{Transit light curves and posterior models for KOI-89.02. The blue filled circles show the observed flux values, and the orange lines show the maximum likelihood model with its standard deviation (shaded region).}
\label{fig:transits_02}
\end{figure*}

\begin{figure*}
\epsscale{1.15}
\plottwo{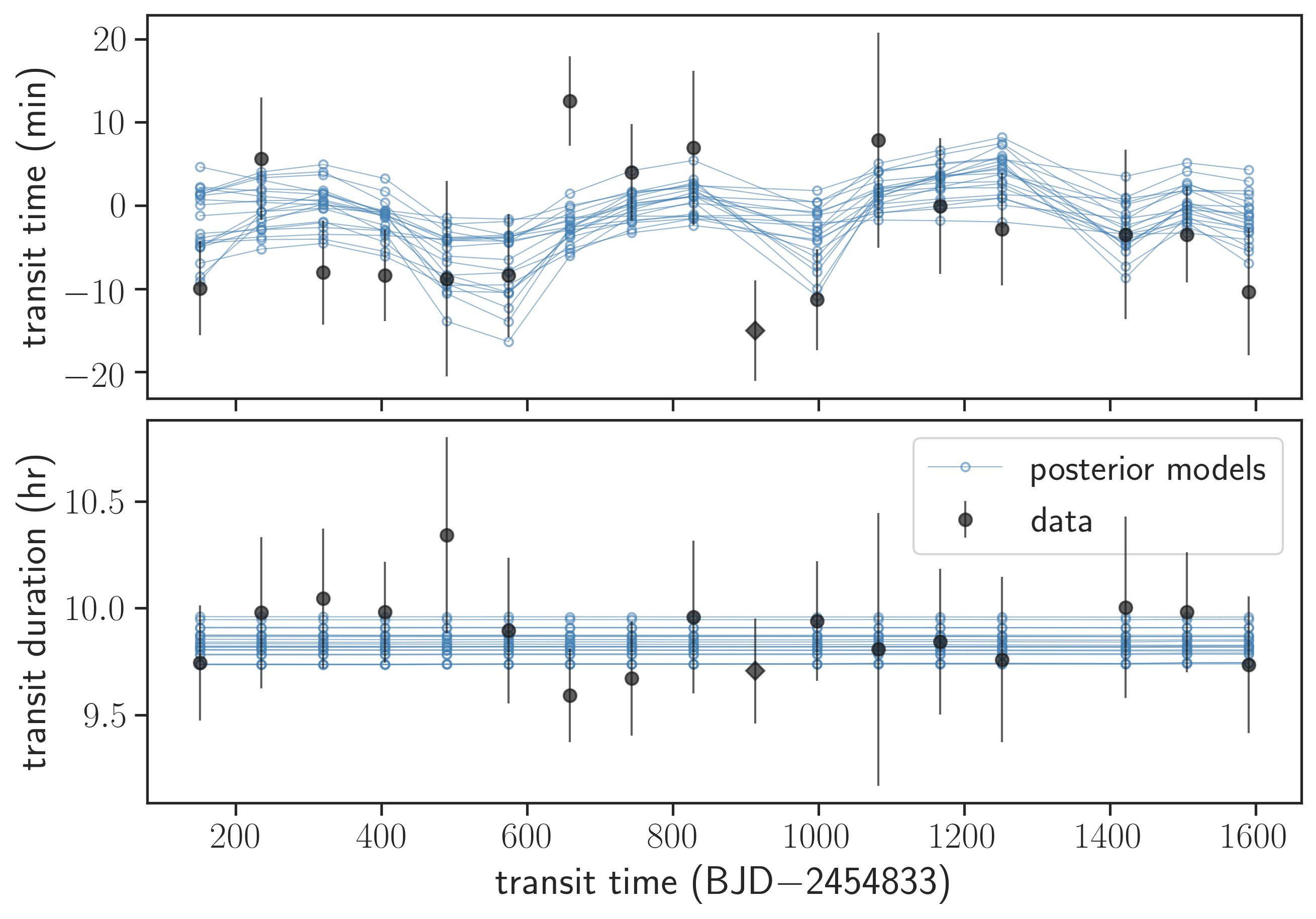}{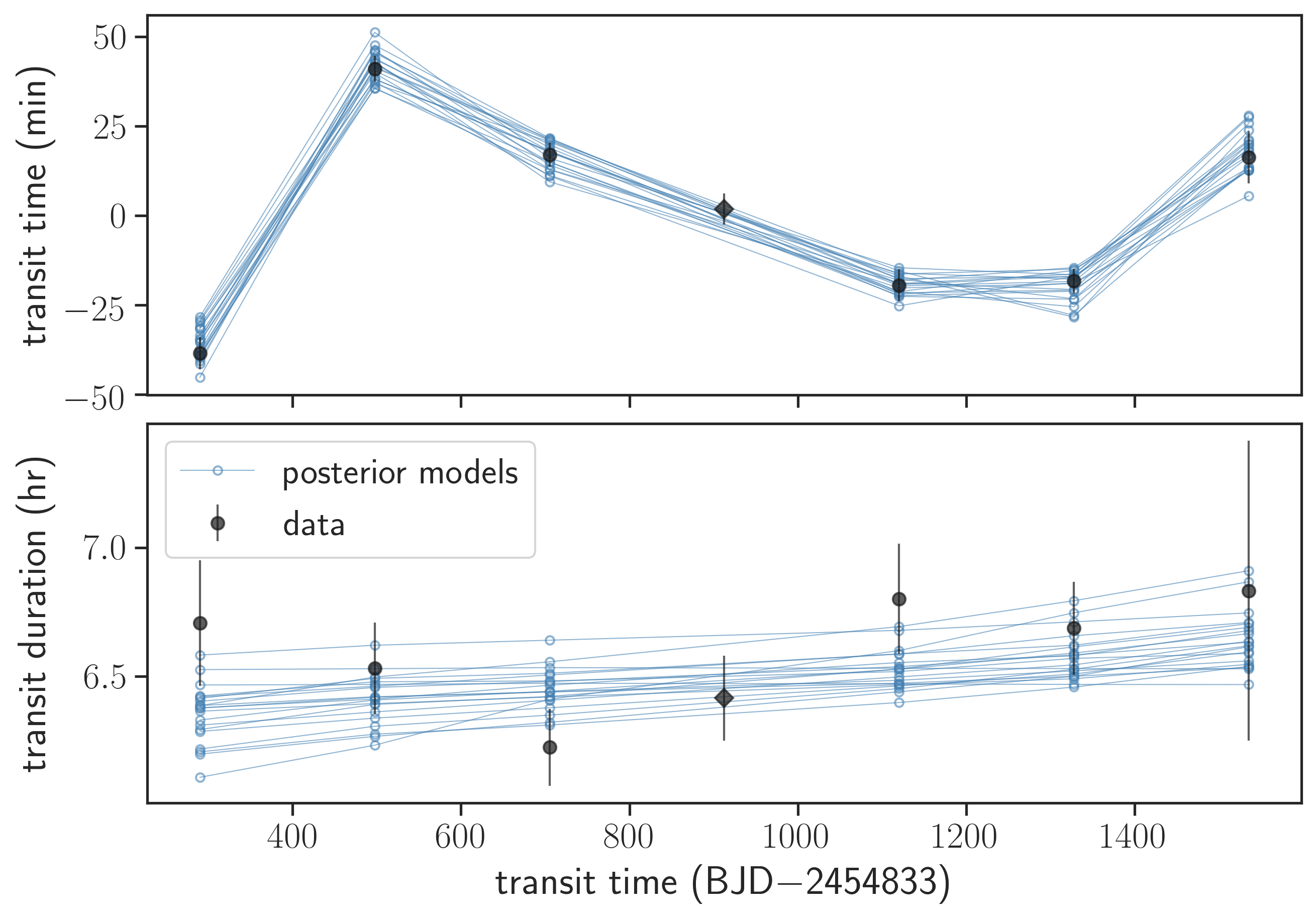}
\caption{Transit times and durations measured by fitting individual transits (gray filled circles with error bars) and posterior models (blue open circles connected with lines). Note that these data are merely meant for illustration; we directly modeled the whole transit light curves (shown in Figure \ref{fig:transits_01} and \ref{fig:transits_02}).  {\it Left} --- KOI-89.01, {\it Right} --- KOI-89.02.}
\label{fig:ttvs}
\end{figure*}

Figures \ref{fig:transits_01} and \ref{fig:transits_02} compare the posterior models with the data for KOI-89.01 and KOI-89.02, respectively. The constraints on the parameters are summarized in Table \ref{tab:koi89}. Here the mass and radius ratios and stellar density derived from the dynamical modeling were converted into physical masses and radii by sampling the stellar mass from a Gaussian with the central value $1.59\,\msun$ and width $0.08\,\msun$ (Section \ref{sec:star}). In Figure \ref{fig:ttvs}, mid-transit times and durations computed for posterior samples of the model parameters are compared with the values measured by fitting individual transits (see Section \ref{ssec:ttv}).
This comparison is to illustrate that the variations in transit times and durations are well captured in our photodynamical modeling, in which we model the whole transit light curves as shown in Figures \ref{fig:transits_01} and \ref{fig:transits_02}. We note that the transit times and durations derived by fitting the simultaneous transits of the two planets (shown as diamonds in this figure), which were not included in the photodynamical modeling, are in reasonable agreement with the posterior predictions conditioned on the other transits. This serves as a sanity check of the modeling. 

We checked for long-term stability of our solution by integrating orbital configurations randomly drawn from our posterior samples. The integrations were performed using the {\sc IAS15} integrator \citep{2015MNRAS.446.1424R} as implemented in the {\tt REBOUND} package \citep{2012A&A...537A.128R}. The stepsize is automatically controlled to maintain numerical errors near machine precision. All the solutions turned out to be dynamically stable at least for $10^7$ inner orbits (i.e., 2.3~Myr) and no sign of instability was found.

From this analysis, we found lower and upper limits on the masses of KOI-89.01 and KOI-89.02, respectively. The former comes from strong TTVs of KOI-89.02 (see top-right panel of Figure \ref{fig:ttvs}), and the latter is due to the weak, if any, TTVs of KOI-89.01 (top-left panel of Figure \ref{fig:ttvs}). The orbital eccentricities are constrained to be low, mainly by the TTV signals and partly by a combination of the transit shape and prior constraint on the mean stellar density.

The orbital misalignment in the sky plane ($\Omega_1$ in Table \ref{tab:koi89}, approximately equivalent to the mutual inclination given that both planets transit) is found to be small from the lack of significant transit duration variations (TDVs), which would otherwise be induced through the nodal orbital precession caused by the planets' mutual gravitational interactions \citep{2002ApJ...564.1019M}. 
We note that the high ($\sim0.9$) impact parameter of KOI-89.02 makes its transit duration particularly sensitive to the perturbations perpendicular to its orbit, and that a weak variation is apparent in Figure \ref{fig:ttvs} even for our solution with a small mutual inclination. See Appendix \ref{sec:misaligned} for more detailed discussion on this mutual inclination constraint.

The weak constraints on mass ratios from this analysis hint that the result can be sensitive to the adopted prior.
Thus, we repeated the analysis with uniform priors on mass ratios whose ranges were selected to be $[0, 200]\,M_\oplus/M_\odot$ and $[0, 50]\,M_\oplus/M_\odot$ for KOI-89.01 and KOI-89.02 respectively, based on the above result and the radii ($\approx 3\,R_\oplus$) of the planets. The range of $\Omega_1$ was also narrowed down to $[-30^\circ, 30^\circ]$ to speed up the convergence. 
The results are summarized in the right part of Table \ref{tab:koi89}.
As expected, the marginal posteriors for the mass ratios changed slightly, although we again recover firm lower and upper mass limits for KOI-89.01 and KOI-89.02 respectively. Additionally, the solution still points to near-circular, well-aligned orbits. We adopt this latter result assuming uniform priors on the mass ratios as the canonical one.

We note that the above solution is consistent with a previous dynamical analysis performed by \citet{2015ApJ...814...67A}. Although their modeling of the light curve shape favored $e_2\gtrsim0.5$, they found that such solutions are not dynamically stable for $10^8\,\mathrm{yr}$ or longer. Instead, they found that the system can be long-term stable if $e_2\lesssim0.35$ and the orbits are co-planar. Here we presented a solution that satisfies this requirement and also explains the transit shapes and TTVs. We will show in Section \ref{sec:gravity_darkening} that this remains the case even if we take into account the possible effect of rapid stellar rotation on the transit shapes.

\subsubsection{Dynamically Unstable Solutions Involving Large Eccentricities and Mutual Orbital Misalignment}\label{sssec:photod_two_eccentric}

The transit duration of KOI-89.02 ($\approx 7\,\mathrm{hr}$) is rather short given its orbital period of $208\,\mathrm{days}$. In the analysis above, we find that this is well explained by a high impact parameter of the planet. On the other hand, the short transit duration alone is also compatible with a low impact parameter transit happening around the periastron of a highly eccentric orbit, as shown in Appendix \ref{ssec:phase}, and this was indeed the class of solutions favored by \citet{2015ApJ...814...67A}.
Such a solution might have been missed due to the too restrictive prior we imposed (i.e., $e_1<0.8$ and $e_2<0.45$). This motivated us to search for solutions where the eccentricity of KOI-89.02 is high. To cover the solution we did not explore, we repeated the photodynamical analysis choosing the prior $e_1 \sim \mathcal{U}(0, 0.9)$ and $e_2 \sim \mathcal{U}(0.45, 0.9)$. The priors on the other parameters were chosen to be the same, and log-uniform priors on the mass ratios were adopted.

From this analysis we found two solutions (i.e., two distinct posterior modes) implying $e_2 \sim 0.8$ and highly inclined ($\sim 120^\circ$) prograde and retrograde orbits.
These solutions equally well match the observed light curves as the low-eccentricity solution, with the difference in the maximum log likelihood being less than unity. 

However, these solutions would be short-lived: the same calculations as performed above show that 90 out of 100 randomly sampled systems become unstable within $10^7$ inner orbits, and nine of the remaining 10 solutions exhibit $>1\%$ variations in the inner semi-major axis at the end of the integration, hinting that the orbits would become unstable on longer timescales. 
Thus we conclude that these solutions are strongly disfavored in light of the $\sim\,\mathrm{Gyr}$ age of the host star.

\begin{deluxetable*}{l@{\hspace{.1cm}}cc@{\hspace{.1cm}}cc@{\hspace{.1cm}}c}[!ht]
\tablecaption{Parameters of the KOI-89 System from Photodynamical Modeling.\label{tab:koi89}}
\tablehead{
\colhead{} & \multicolumn{2}{c}{log-flat prior on mass ratios} & \multicolumn{2}{c}{\textbf{flat prior on mass ratios}}\\
\colhead{} & \colhead{MAP \& $68\%$ HDI} & \colhead{$95\%$ HDI} & \colhead{MAP \& $68\%$ HDI} & \colhead{$95\%$ HDI} & \colhead{prior} 
}
\startdata
\textit{(host star)}\\
mean density $\rho_\star$ ($\mathrm{g\,cm^{-3}}$) & $0.50^{+0.06}_{-0.06}$ & $[0.40, 0.61]$ &   $0.56^{+0.07}_{-0.07}$ & $[0.44, 0.68]$ & $\g(0.53, 0.06)$  \\
limb-darkening coefficient $(u_1+u_2)^2$ & $0.20^{+0.08}_{-0.06}$ & $[0.09, 0.37]$ &   $0.22^{+0.08}_{-0.07}$ & $[0.10, 0.40]$ & $\u(0,1)$  \\
limb-darkening coefficient $u_1/2(u_1+u_2)$ & $0.3^{+0.2}_{-0.2}$ & $[0.0, 0.7]$ &   $0.3^{+0.2}_{-0.2}$ & $[0.0, 0.7]$ &  $\u(0,1)$ \\
radius $R_\star$ ($R_\odot$) & $1.63^{+0.07}_{-0.06}$ & $[1.52, 1.78]$ &   $1.57^{+0.08}_{-0.06}$ & $[1.47, 1.73]$ & \nodata \\
\textit{(KOI-89.01)}\\
mass ratio $m_1/M_\star$ ($M_\oplus/M_\odot$) & $17.2^{+14.7}_{-4.9}$ & $[10.1, 52.4]$ &   $35.7^{+64.2}_{-16.5}$ & $[16.4, 174.2]$ &  $\logu(0.5,100)$ / $\u(0, 200)$\\
orbital period $P_1$ (days) & $84.6873^{+0.0003}_{-0.0005}$ & $[84.6861, 84.6878]$ &   $84.6866^{+0.0007}_{-0.0010}$ & $[84.6845, 84.6877]$ & $\u(84.58,84.78)$ \\
eccentricity $e_1$ & $0.12^{+0.05}_{-0.03}$ & $[0.07, 0.21]$ &   $0.042^{+0.046}_{-0.008}$ & $[0.031, 0.152]$ &  $\u(0, 0.8)$ / $\u(0, 0.4)$\\
argument of periastron $\omega_1$ (deg) & $57.0^{+7.9}_{-7.9}$ & $[42.2, 78.2]$ &   $51.9^{+7.1}_{-11.3}$ & $[32.1, 66.1]$ &  $\u(-180,180)$ \\
impact parameter $b_1$ ($R_\star$) & $0.1^{+0.1}_{-0.2}$ & $[-0.3, 0.3]$ &   $-0.31^{+0.57}_{-0.06}$ & $[-0.47, 0.44]$ & $\u(-1,1)$\\
longitude of ascending node $\Omega_1$ (deg) & $-3.9^{+2.8}_{-5.2}$ & $[-16.1, 1.3]$ &   $-1.2^{+1.1}_{-2.0}$ & $[-8.5, 0.7]$ &  $\u(-180,180)$ / $\u(-40,40)$\\
time of inferior conjunction $t_{\rm c,1}$ & $150.574^{+0.002}_{-0.002}$ & $[150.571, 150.577]$ &   $150.573^{+0.002}_{-0.002}$ & $[150.569, 150.577]$ & $\u(150.47,150.67)$   \\
radius ratio $r_1/R_\star$ & $0.0175^{+0.0001}_{-0.0002}$ & $[0.0172, 0.0178]$ &   $0.0175^{+0.0002}_{-0.0002}$ & $[0.0172, 0.0179]$ & $\logu(0.001,0.1)$   \\
mass $m_1$ ($M_\oplus$) & $27.9^{+23.3}_{-8.2}$ & $[15.7, 83.6]$ &   $53.9^{+104.6}_{-23.8}$ & $[26.4, 278.6]$ &   \nodata \\
radius $r_1$ ($R_\oplus$) & $3.1^{+0.1}_{-0.1}$ & $[2.9, 3.4]$ &   $3.0^{+0.2}_{-0.1}$ & $[2.8, 3.3]$ &  \nodata \\
mean density $\rho_1$ ($\mathrm{g\,cm^{-3}}$) & $4.3^{+4.9}_{-1.4}$ & $[2.3, 16.9]$ &   $9.3^{+22.3}_{-4.6}$ & $[3.6, 57.2]$ & \nodata \\
\textit{(KOI-89.02)}\\
mass ratio $m_2/M_\star$ ($M_\oplus/M_\odot$) & $0.5004^{+1.6845}_{-0.0004}$ & $[0.5000, 5.7624]$ &   $4.3^{+6.3}_{-3.5}$ & $[0.0, 18.3]$ & $\logu(0.5,100)$ / $\u(0, 100)$   \\
orbital period $P_2$ (days) & $207.620^{+0.008}_{-0.008}$ & $[207.605, 207.635]$ &   $207.62^{+0.03}_{-0.01}$ & $[207.61, 207.69]$ & $\logu(207.38	, 207.78)$   \\
eccentricity $e_2$ & $0.24^{+0.02}_{-0.06}$ & $[0.14, 0.29]$ &   $0.11^{+0.05}_{-0.02}$ & $[0.09, 0.23]$ & $\u(0, 0.45)$  \\
argument of periastron $\omega_2$ (deg) & $12.8^{+4.5}_{-7.6}$ & $[-3.3, 21.4]$ &   $14.9^{+3.2}_{-3.2}$ & $[7.3, 21.5]$ & $\u(-180,180)$  \\
impact parameter $b_2$ ($R_\star$) & $0.896^{+0.010}_{-0.008}$ & $[0.881, 0.914]$ &   $0.890^{+0.011}_{-0.008}$ & $[0.874, 0.910]$ & $\u(0,1)$  \\
time of inferior conjunction $t_{\rm c,2}$ & $289.863^{+0.008}_{-0.005}$ & $[289.854, 289.881]$ &   $289.87^{+0.03}_{-0.01}$ & $[289.86, 289.94]$ & $\u(289.66,290.06)$  \\
radius ratio $r_2/R_\star$ & $0.0232^{+0.0004}_{-0.0003}$ & $[0.0226, 0.0239]$ &   $0.0232^{+0.0004}_{-0.0003}$ & $[0.0225, 0.0239]$ &  $\logu(0.001,0.1)$ \\
mass $m_2$ ($M_\oplus$) & $1.1^{+2.4}_{-0.3}$ & $[0.7, 9.2]$ &   $6.0^{+10.8}_{-4.9}$ & $[0.0, 29.3]$ & \nodata   \\
radius $r_2$ ($R_\oplus$)& $4.1^{+0.2}_{-0.2}$ & $[3.8, 4.6]$ &   $4.0^{+0.2}_{-0.2}$ & $[3.7, 4.4]$ & \nodata  \\
\vspace{0.1cm}
mean density $\rho_2$ ($\mathrm{g\,cm^{-3}}$) & $0.08^{+0.18}_{-0.03}$ & $[0.05, 0.77]$ &   $0.4^{+1.1}_{-0.3}$ & $[0.0, 2.6]$ & \nodata \\
mutual orbital inclination $i_{12}$ (deg) & $3.8^{+4.1}_{-3.2}$ & $[0.3, 15.4]$ &   $1.1^{+2.0}_{-0.7}$ & $[0.1, 7.9]$ &  \nodata \\
\textit{(noise model)}\\
$\ln \sigma_{\rm jit}$ & $-10.7^{+0.1}_{-0.2}$ & $[-11.1, -10.5]$ &   $-10.7^{+0.1}_{-0.2}$ & $[-11.2, -10.4]$ & $\u(-13,-7)$  \\
$\ln \alpha$ & $-10.9^{+0.1}_{-0.1}$ & $[-11.2, -10.6]$ &   $-10.9^{+0.1}_{-0.2}$ & $[-11.2, -10.6]$ &  $\u(-13,-7)$ \\
$\ln \rho$ (days) & $-2.6^{+0.3}_{-0.3}$ & $[-3.3, -2.1]$ &   $-2.6^{+0.3}_{-0.4}$ & $[-3.4, -2.1]$ & $\u(-5,1)$   \\
\enddata
\tablecomments{Values listed here report the maximum a posteriori (MAP) and 68\%/95\% highest density intervals (HDIs) of the marginal posteriors. Values without specified priors are the parameters that were not fitted directly, and are derived assuming a Gaussian distributed $M_\star=1.59\pm0.08\,M_\odot$ when necessary. Longitude of the ascending node of KOI-89.02 is fixed to be zero, and arguments of the periastron are referred to the sky plane.
Priors --- $\mathcal{N}(\mu,\sigma)$ means the gaussian PDF centered on $\mu$ and width $\sigma$; $\u(a,b)$ is the uniform PDF between $a$ and $b$; $\logu(a,b)$ is the log-uniform PDF between $a$ and $b$. Dots indicate the parameters that were not directly sampled but were derived from the samples of the ``fitted" parameters.}
\end{deluxetable*}

\subsection{Results: Three-planet Model}\label{ssec:photod_three}

In Section \ref{ssec:photod_two}, we showed that the two-planet model well explains the transit light curves of the two planets including TTVs. This is the simplest model given the current data, and there is no strong need to add another planet. Nevertheless, here we explore three-planet models motivated by an unusually large mass ($\gtrsim20\,M_\oplus$) estimated for the inner KOI-89.01 given its size ($\approx3\,R_\oplus$), as we will discuss in more detail later (Section \ref{ssec:discussion_mass}). This is a solid lower limit imposed by the large TTVs exhibited by KOI-89.02, whose orbit is not particularly close to that of KOI-89.01, nor to any low-order mean-motion resonances with it. 

Here we fixed $\log_{10}(m_1/M_\star)=0.77$ based on the \citet{2017ApJ...834...17C} mass--radius relation as a ``reasonable" value for $m_1$, and searched for solutions where the third planet (instead of KOI-89.01) may explain the TTVs of KOI-89.02. Here we assume that the three planets have mostly aligned orbits by fixing the longitudes of ascending nodes to be the same value and $\cos i_3$ to be zero, in the spirit of searching for solutions more typical of {\it Kepler} multi-planet systems.

Even under this simplification, given the sparse TTV data and the several new degrees of freedom introduced by the third planet, we expect that the solution will be complicated and multimodal. We therefore divided the prior range for $P_3$ into $24$ equally-spaced log-intervals between $1.05\,P_2$ and $4\,P_2$, obtained posterior samples separately using the same {\tt MultiNest} fitting, and combined them by weighing the evidence value for each interval. Mathematically, the normalized ``partial" prior for the $i$th bin $p_i(P_3)$ (i.e., log-flat in the $i$th bin and zero otherwise) and the ``full" normalized prior are related by $p(P_3)=\sum_i p_i(P_3)/N$ with $N$ being the number of $P_3$ bins. Therefore,
\begin{align}
	\notag
	p(\bm{\theta},P_3|\bm{d}) 
	&= {1\over N}\sum_i {p(\bm{d}|\bm{\theta}, P_3)\over p(\bm{d})}\,p(\bm{\theta})\,p_i(P_3)\\
	&= {1\over N}\sum_i {p_i(\bm{d})\over p(\bm{d})}\, p_i(\bm{\theta}, P_3|\bm{d}),
\end{align}
where $\bm{\theta}$ is the set of parameters other than $P_3$, $p_i(\bm{\theta}, P_3|\bm{d})$ is the posterior conditioned on the partial prior $p(\bm{\theta})p_i(P_3)$, and $p_i(\bm{d})$ is the corresponding evidence.
Although the reliability of this ``combined" posterior sample may be limited due to the limited accuracy of evidence evaluation, we believe that the sample is sufficient for our current purpose to capture the plausible range of three-planet solutions.

We find that the three-planet model can explain the data equally well as the two-planet model. The maximum likelihood value is slightly higher, but not high enough to favor the three-planet hypothesis over the two-planet one, considering the increased model complexity (e.g., in terms of the Bayesian information criterion). The constraints for the selected parameters are summarized in Table \ref{tab:koi89_3}, although we note that $m_3$ and $P_3$ have multimodal marginal posteriors and the statistics here do not fully capture the complex nature of the solutions. We found significantly non-zero $m_3$, which is consistent with our inference in the two-planet model that ``normal" values of $m_1$ are insufficient to explain the strong TTVs of KOI-89.02. On the other hand, we found relatively well-constrained posteriors for the parameters of the inner two planets, including the mass of KOI-89.02 and eccentricities of both KOI-89.01 and KOI-89.02. In this model two planets both have $10$--$20\,M_\oplus$ (for $M_\star=1.59\,M_\odot$) and low eccentricities.

\begin{deluxetable}{l@{\hspace{.5cm}}cc@{\hspace{.5cm}}c}[!ht]
\tablecaption{Masses and eccentricities from the three-planet photodynamical model.\label{tab:koi89_3}}
\tablehead{
 & \colhead{68\% HDI} & \colhead{95\% HDI} & \colhead{prior}
}
\startdata
$m_1/M_\star$ ($M_\oplus/M_\star$) & \multicolumn{2}{c}{$5.89$} & fixed\\
$m_2/M_\star$  ($M_\oplus/M_\star$)&  $[7.6, 14.7]$ &  $[5.6, 20.3]$ & $\u(0, 50)$\\
$m_3/M_\star$ ($M_\oplus/M_\star$) & $[43, 93]$ &  $[10, 94]$ & $\u(0, 100)$\\
$e_1$ & $[0.04, 0.12]$ & $[0.03, 0.19]$ & $\u(0, 0.4)$\\
$e_2$ & $[0.00, 0.02]$ & $[0.00, 0.06]$ & $\u(0, 0.4)$\\
$e_3$ & $[0.01, 0.05]$ & $[0.00, 0.10]$ & $\u(0, 0.4)$\\
$P_3$ (days) & $[478, 548]$ & $[351, 739]$ & $\logu(218, 830)$\\ 
$m_2$ ($M_\oplus$)  &  $[12.2, 23.6]$ &  $[8.8, 32.5]$  & \nodata \\
$m_3$ ($M_\oplus$)  &  $[65, 149]$ &  $[15, 152]$ & \nodata  \\
\enddata
\tablecomments{Values listed here report the 68\%/95\% highest density intervals (HDIs) of the marginal posteriors derived using the method described in Section \ref{ssec:photod_three}. Note that marginal posteriors for the mass and period of the third planet are multimodal and these parameters are poorly constrained.  The physical masses are derived assuming a Gaussian distributed $M_\star=1.59\pm0.08\,M_\odot$ when necessary. Priors --- $\mathcal{N}(\mu,\sigma)$ means the gaussian PDF centered on $\mu$ and width $\sigma$; $\u(a,b)$ is the uniform PDF between $a$ and $b$; $\logu(a,b)$ is the log-uniform PDF between $a$ and $b$. Dots indicate the parameters that were not directly sampled but were derived from the samples of the ``fitted" parameters.}
\end{deluxetable}

\subsection{Conclusions}

The findings from our photodynamical analysis are summarized as follows.
\begin{itemize}
\item Significant TTVs of KOI-89.02 suggest either an unusually large mass ($\gtrsim 20\,M_\oplus$) for KOI-89.01 given its radius, or the presence of an undetected third planet. 
\item In both of the two-planet and three-planet models, the mass of KOI-89.02 is in the planetary range, and the orbits of KOI-89.01 and KOI-89.02 have low eccentricities.\footnote{The conclusion should remain unchanged even if the TTVs of KOI-89.02 is caused by a third planet in between KOI-89.01 and KOI-89.02, in which case there is even smaller room for solutions involving larger masses and eccentricities.}
In the two-planet model, a low mutual inclination is also required by the data.
\end{itemize}

The above analysis also demonstrates an advantage of full photodynamical modeling over a two-step approach in which the TTVs and TDVs are derived first and then fitted with a dynamical model. 
The latter approach usually employs additional constraint on the mean impact parameter or the in-transit velocity (i.e., combination of the mean stellar density, eccentricity, and argument of periastron) derived from the transit shape, to solve the degeneracy between eccentricities and inclinations \citep[e.g.,][]{2012Natur.487..449S}.
This works well if the impact parameter is well constrained, but for KOI-89.02 the eccentricity and impact parameter are strongly degenerate as shown explicitly in Appendix \ref{ssec:phase}. In such a case, the result of the dynamical fit can depend sensitively on how those constraints from transit shapes are incorporated, and the issue could become more serious if the correlate noise in the light curve (which we modeled here) significantly affects those constraints.
For example, if one adopts the mean ($\approx0.6$) of the marginal posterior of the KOI-89.02's transit impact parameter (Appendix \ref{ssec:phase}) and fits TTVs and TDVs, the modeling may erroneously favor high-eccentricity and mutually inclined solutions as we explored in Section \ref{sssec:photod_two_eccentric}. For KOI-89, we were able to exclude these solutions based on stability argument in any case, but that may not always be possible. 
See also \citet{2020AJ....159..223D} for an in-depth study of related issues.

\section{Analysis of Gravity-Darkened Transit Light Curves}\label{sec:gravity_darkening}

The star KOI-89 has a high projected rotation velocity $v\sin i_\star$ of $\approx 80\,\mathrm{km/s}$ (Section \ref{sec:star}). The rapid rotation results in the distorted stellar surface as well as associated inhomogeneity in the surface brightness distribution, the effect known as gravity darkening \citep{1924MNRAS..84..665V}. The non-axisymmetric surface brightness distribution induced by the gravity darkening causes distortion of the transit light curve in such a way that depends on the stellar obliquity \citep{2009ApJ...705..683B}. \citet{2015ApJ...814...67A} argued that the transit of KOI-89.02 does exhibit gravity-darkening distortions and concluded that the star's spin axis is tilted by $\approx 70^\circ$ from the orbital axes of the planets. The analysis used the stacked and binned transit light curves of the two planets, which involved a correction for the significant TTVs of KOI-89.02 and an inference of the unknown orbital eccentricity. We now revisit this analysis with our own TTV corrections and the associated prior knowledge on the eccentricities. 

We analyze the stacked, long-cadence transit light curves of the two planets simultaneously. Given the significant TTVs of KOI-89.02, the stacking was performed by shifting each transit so that its central time becomes zero using the transit times derived by fitting each transit in Section \ref{ssec:photod_data}. 
We did not correct for possible duration variations, which we did not detect significantly. Unlike in \citet{2015ApJ...814...67A}, we chose {\it not} to bin the stacked light curves, because binning the long-cadence light curve may introduce additional shape distortion that is difficult to model and because the number of data points is small due to the long orbital period. 

We modeled the gravity-darkened transit light curve as described in \citet{2015ApJ...805...28M}, which is based on the formulation in \citet{2009ApJ...705..683B}. The model involves the following parameters: mass, mean density, spin inclination measured from the line of sight $i_\star$, projected rotation velocity, effective temperature at the pole, gravity-darkening exponent, limb-darkening coefficients for the star; and time of inferior conjunction, orbital period, eccentricity, argument of periastron, impact parameter, ratio of the planet radius to the stellar equatorial radius, and sky-projected obliquity $\lambda$ for each of the planets. Here the stellar effective temperature was fixed to be 7540~K, and the limb-darkening coefficients were parameterized as in \citet{2013MNRAS.435.2152K} and treated as free parameters. Following \citet{2020ApJ...888...63A}, the gravity darkening coefficient was chosen to be a deterministic function of the stellar oblateness (which in our model depends on the mass, mean density, projected rotation velocity, and inclination of the star) using the theoretical calculation in \citet{2011A&A...533A..43E}; although the resulting values turned out to be fairly close to the standard value of 0.25 because the stellar oblateness was found to be $\mathcal{O}(1\%)$. We modeled the noise as a Gaussian process with the same Mat\'ern-3/2 kernel and the white-noise term as in Equation \ref{eq:kernel}, and marginalized over the noise parameters. 

We fit the two transit light curves simultaneously, and obtained posterior samples of the model parameters  using {\tt MultiNest} \citep{2008MNRAS.384..449F, 2009MNRAS.398.1601F, 2013arXiv1306.2144F} and its python interface {\tt PyMultiNest} \citep{2014A&A...564A.125B}. We assumed separable, uniform or log-uniform priors for the model parameters (see Table \ref{tab:gdfit}) except for the following parameters.
The prior for the mean density was chosen to be a Gaussian with mean $0.53\,\mathrm{g\,cm^{-3}}$ and dispersion $0.06\,\mathrm{g\,cm^{-3}}$. For projected rotation velocity, we assumed a Gaussian with mean $85\,\mathrm{km/s}$ and dispersion $5\,\mathrm{km/s}$. The choice is motivated by our derived value of $v\sin i_\star=80\pm3\,\mathrm{km/s}$ and the value of $90\,\mathrm{km/s}$, which was adopted in \citet{2015ApJ...814...67A} based on the data from the {\it Kepler} Community Follow-up Observing Program.
The prior on the eccentricities and arguments of periastron of the two planets was defined by the kernel density estimation of the joint posterior samples for the four parameters from the photodynamical modeling in Section \ref{sec:photod}. We did so both using the results from the two-planet model (Section \ref{ssec:photod_two}) and the three-planet model (Section \ref{ssec:photod_three}) and performed separate analyses. 

\begin{figure}
	\epsscale{1.15}
	\plotone{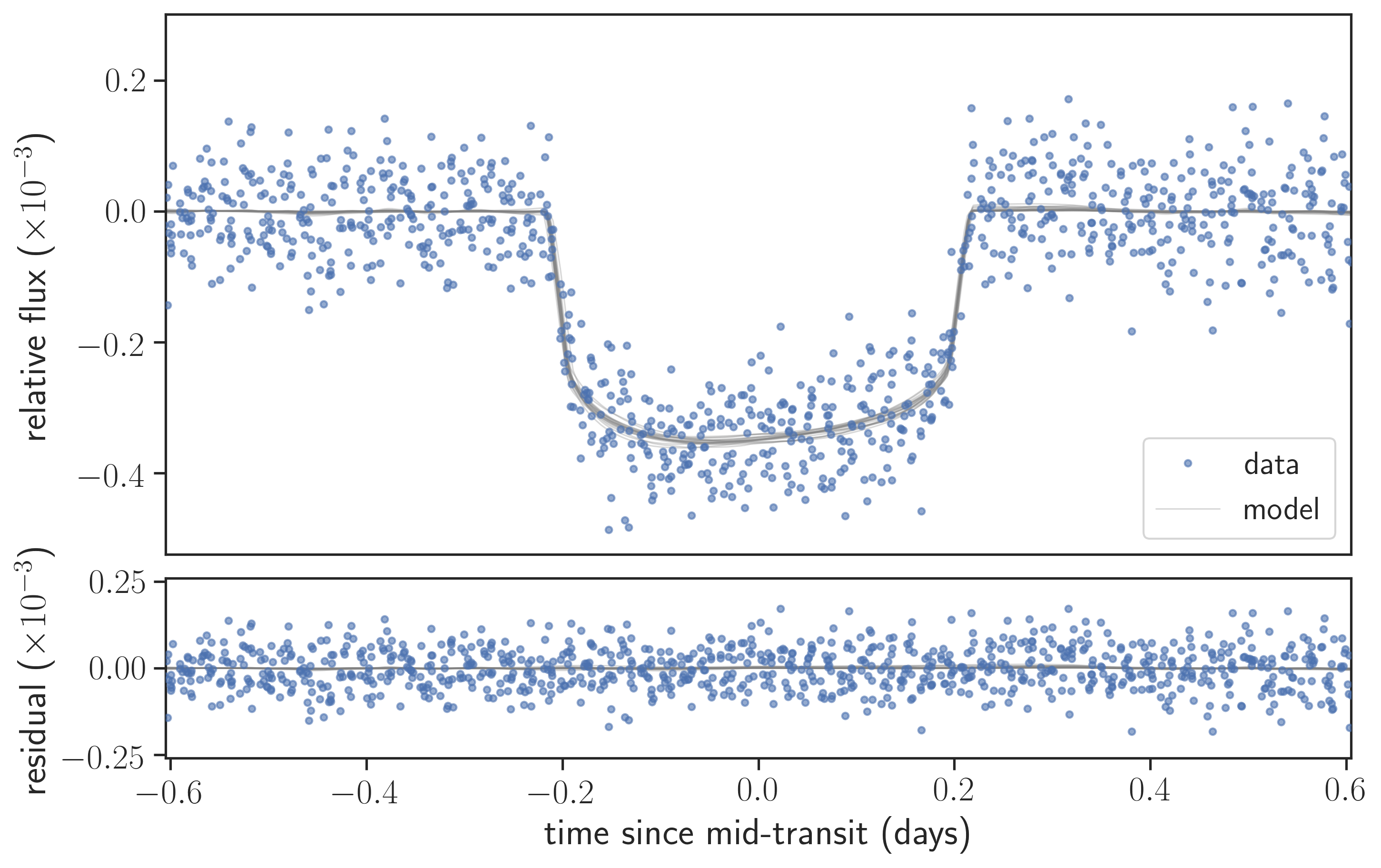}
	\plotone{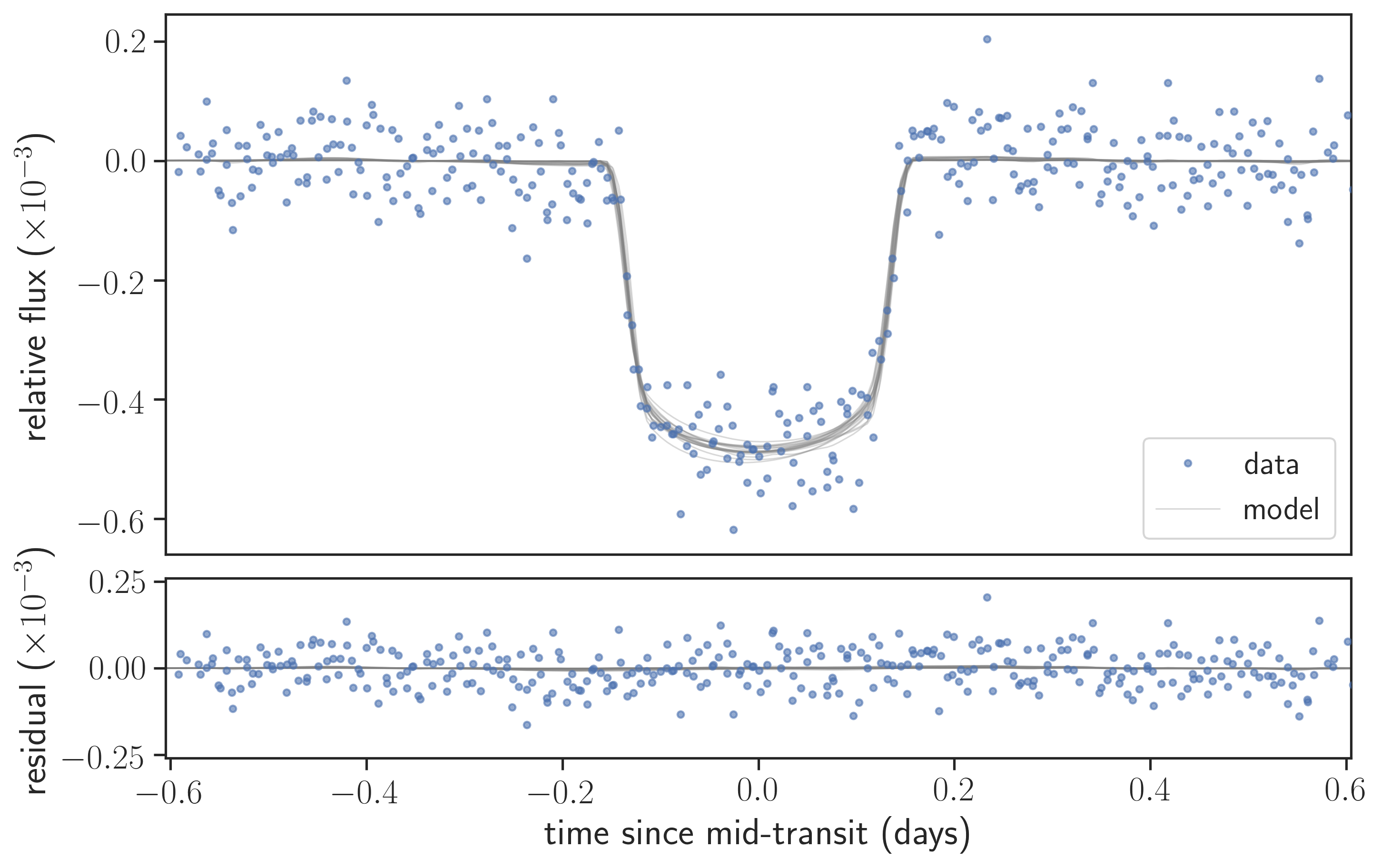}
	\caption{Modeling of the stacked transit light curves of KOI-89.01 (top) and KOI-89.02 (bottom). Blue dots show all the long-cadence data points. Gray solid lines are the gravity-darkened models computed for 20 parameter sets drawn from the posterior. The top panels show the flux data and model relative to one. The bottom panels show them after subtracting the maximum-likelihood transit model.}
	\label{fig:gdfit}
\end{figure}

\begin{figure}
	\epsscale{1.15}
	\plotone{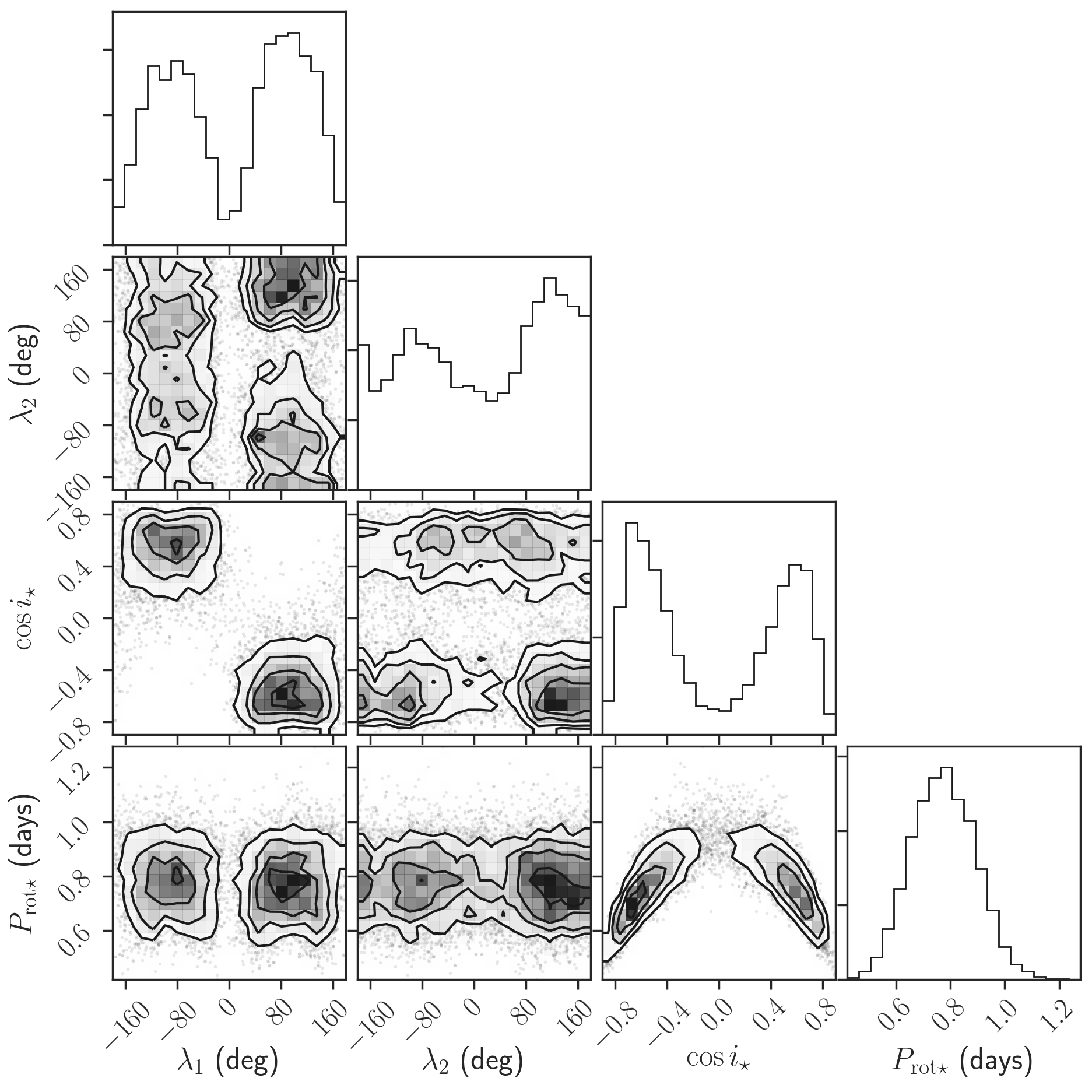}
	\plotone{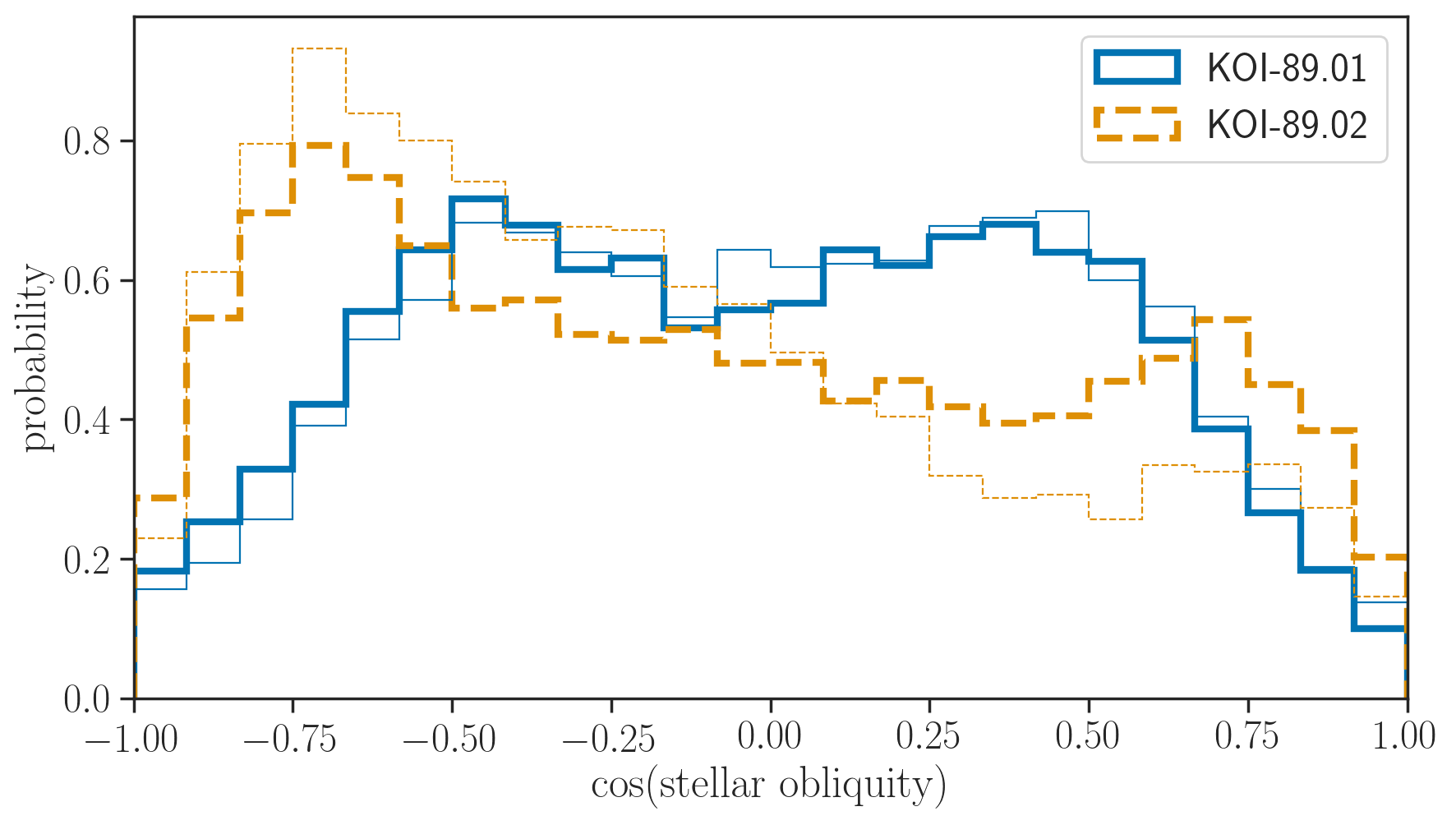}
	\caption{{\it Top} --- Corner plot for the obliquity related parameters from the gravity-darkened fit. Here the prior from two-planet photodynamical model is adopted. {\it Bottom} --- Marginal posterior distributions for the cosine of the stellar obliquity with respect to the orbits of KOI-89.01 (solid) and KOI-89.02 (dashed). Thicker lines are the results based on the eccentricity prior from the two-planet photodynamical model (Section \ref{ssec:photod_two}) and thinner lines are based on the prior from the three-planet model (Section \ref{ssec:photod_three}).}
	\label{fig:obliquity}
\end{figure}

Figure \ref{fig:gdfit} shows the stacked light curves with the posterior samples for the model light curves. Table \ref{tab:gdfit} summarizes the constraints from marginal posteriors. Figure \ref{fig:obliquity} (top) shows a corner plot for the posterior samples of obliquity-related parameters, and Figure \ref{fig:obliquity} (bottom) shows the marginal posterior distributions for the stellar obliquity relative to the orbits of the two planets. These results show that stellar obliquity is poorly constrained, both for the priors based on the two-planet and three-planet photodynamical models. This essentially means that the gravity darkening effect is not significantly detected in the light curves: in our solutions, gravity darkening causes $\mathcal{O}(1\%)$ change in the flux loss, which translates into $\sim 5\times 10^{-6}$ in terms of the relative flux even for KOI-89.02 with the transit depth of $5\times 10^{-4}$. This is much smaller than the observed noise level as seen, for example, in the bottom panels of Figure \ref{fig:gdfit}.

We also find that planet-to-star radius ratios and transit impact parameters from the gravity-darkened model agree with those from photodynamical modeling within 2--3$\sigma$.
This agreement is satisfactory given that the photodynamical model ignores $\mathcal{O}(1\%)$ distortion of the star: although we did not detect the gravity darkening effect on the surface brightness distribution significantly as discussed above, the model still incorporates geometric distortion (oblateness) of the star that does affect the estimated impact parameter and radius ratio.
This systematic error is much smaller than the precision of the physical parameters of the system (including planet mass and radius) which is limited by the precision of the stellar mass determination. Thus the photodynamical modeling in Section \ref{sec:photod} without taking into account gravity darkening is justified.

\begin{deluxetable*}{l@{\hspace{.1cm}}cc@{\hspace{.1cm}}cc@{\hspace{.3cm}}c}[!ht]
\tablecaption{Parameters from the Gravity-darkened Fit to the Stacked Transit Light Curves.\label{tab:gdfit}}
\tablehead{
\colhead{} & \multicolumn{2}{c}{prior from two-planet model} & \multicolumn{2}{c}{prior from three-planet model}\\
\colhead{} & \colhead{MAP \& $68\%$ HDI} & \colhead{$95\%$ HDI} & \colhead{MAP \& $68\%$ HDI} & \colhead{$95\%$ HDI} & \colhead{prior} 
}
\startdata
\textit{(stellar parameters)}\\
limb-darkening coefficient $(u_1+u_2)^2$ & $0.27^{+0.08}_{-0.08}$ & $[0.13, 0.43]$ &   $0.26^{+0.08}_{-0.06}$ & $[0.14, 0.42]$ & $\mathcal{U}(0,1)$ \\
limb-darkening coefficient $u_1/2(u_1+u_2)$ & $0.3^{+0.1}_{-0.2}$ & $[0.0, 0.6]$ &   $0.3^{+0.2}_{-0.1}$ & $[0.0, 0.5]$ & $\mathcal{U}(0,1)$  \\
mean density $\rho_\star$ ($\mathrm{g\,cm^{-3}}$) & $0.60^{+0.04}_{-0.04}$ & $[0.51, 0.67]$ &   $0.59^{+0.04}_{-0.04}$ & $[0.52, 0.67]$ & $\mathcal{N}(0.53,0.06)$   \\
mass $M_\star$ ($M_\odot$) & $1.6^{+0.1}_{-0.1}$ & $[1.4, 1.8]$ &   $1.59^{+0.09}_{-0.09}$ & $[1.39, 1.79]$ & $\mathcal{N}(1.59,0.08)$  \\
projected rotation velocity $v\sin i_\star$ ($\mathrm{km\,s^{-1}}$) & $84.4^{+6.2}_{-6.0}$ & $[71.4, 96.4]$ &   $83.9^{+5.7}_{-5.6}$ & $[72.3, 95.7]$ & $\mathcal{N}(85,5)$  \\
cosine of spin inclination $\cos i_\star$ & $-0.6^{+1.1}_{-0.1}$ & $[-0.8, 0.8]$ &   $-0.6^{+0.5}_{-0.2}$ & $[-0.8, 0.7]$ & $\mathcal{U}(-1,1)$   \\
rotation period $P_\mathrm{rot\star}$ (days) & $0.8^{+0.1}_{-0.1}$ & $[0.5, 1.0]$ &   $0.8^{+0.1}_{-0.1}$ & $[0.6, 1.0]$ & \nodata \\
\textit{(KOI-89.01)}\\
time of inferior conjunction $t_0$(days) & $0.000^{+0.001}_{-0.001}$ & $[-0.002, 0.003]$ &   $0.0004^{+0.0010}_{-0.0009}$ & $[-0.0014, 0.0024]$ & $\mathcal{U}(-0.01,0.01)$   \\
planet-to-star radius ratio $r/R_\star$& $0.0172^{+0.0001}_{-0.0002}$ & $[0.0170, 0.0176]$ &   $0.0172^{+0.0001}_{-0.0002}$ & $[0.0169, 0.0175]$ & $\mathcal{U}_{\rm log}(0.005,0.05)$  \\
eccentricity $e$ & $0.05^{+0.03}_{-0.01}$ & $[0.03, 0.12]$ &   $0.09^{+0.04}_{-0.03}$ & $[0.04, 0.19]$ &  photodynamical \\
argument of periastron $\omega$ (deg) & $49.8^{+6.6}_{-8.8}$ & $[34.8, 63.2]$ &   $154.9^{+13.6}_{-21.6}$ & $[88.0, 178.1]$ & modeling  \\
impact parameter $b$ ($R_\star$) & $-0.14^{+0.31}_{-0.07}$ & $[-0.29, 0.29]$ &   $-0.07^{+0.23}_{-0.06}$ & $[-0.24, 0.25]$ & $\mathcal{U}(-1,1)$  \\
sky-projected obliquity $\lambda$ (deg) & $78.5^{+51.4}_{-172.6}$ & $[-151.3, 160.1]$ &   $75.3^{+85.4}_{-56.9}$ & $[-136.7, 160.6]$ & $\mathcal{U}(-180,180)$  \\
\textit{(KOI-89.02)}\\
time of inferior conjunction $t_0$ (days) & $0.004^{+0.001}_{-0.005}$ & $[-0.005, 0.006]$ &   $0.003^{+0.002}_{-0.004}$ & $[-0.005, 0.006]$ & $\mathcal{U}(-0.01,0.01)$  \\
planet-to-star radius ratio $r/R_\star$ & $0.0232^{+0.0003}_{-0.0004}$ & $[0.0223, 0.0238]$ &   $0.0233^{+0.0003}_{-0.0004}$ & $[0.0225, 0.0239]$ & $\mathcal{U}_{\rm log}(0.005,0.05)$  \\
eccentricity $e$ & $0.12^{+0.03}_{-0.02}$ & $[0.09, 0.19]$ &   $0.03^{+0.02}_{-0.02}$ & $[0.00, 0.07]$ &  photodynamical  \\
argument of periastron $\omega$ (deg) & $15.1^{+2.7}_{-2.6}$ & $[9.5, 20.4]$ &   $99.5^{+69.5}_{-76.1}$ & $[-153.3, 179.9]$ & modeling  \\
impact parameter $b$ ($R_\star$) & $0.869^{+0.007}_{-0.006}$ & $[0.858, 0.883]$ &   $0.878^{+0.009}_{-0.009}$ & $[0.859, 0.894]$ & $\mathcal{U}(0,1)$  \\
sky-projected obliquity $\lambda$ (deg)& $111.0^{+68.5}_{-171.8}$ & $[-163.7, 178.7]$ &   $111.7^{+68.2}_{-131.9}$ & $[-155.2, 178.8]$ &  $\mathcal{U}(-180,180)$ \\
\textit{(noise parameters)}\\
$\ln \sigma_{\rm jit}$ & $-10.51^{+0.07}_{-0.09}$ & $[-10.68, -10.36]$ &   $-10.51^{+0.06}_{-0.09}$ & $[-10.68, -10.37]$ & $\mathcal{U}(-13,-7)$  \\
$\ln \alpha$ & $-12.3^{+0.5}_{-0.5}$ & $[-13.0, -11.5]$ &   $-12.1^{+0.4}_{-0.5}$ & $[-13.0, -11.5]$ &  $\mathcal{U}(-13,-7)$ \\
$\ln \rho$ (days) & $-4.5^{+1.7}_{-0.5}$ & $[-5.0, -1.0]$ &   $-4.2^{+1.2}_{-0.8}$ & $[-5.0, -1.6]$ & $\mathcal{U}(-5,1)$  \\
\enddata
\tablecomments{Values listed here report the maximum a posteriori (MAP) and 68\%/95\% highest density intervals (HDIs) of the marginal posteriors. Priors --- $\mathcal{N}(\mu,\sigma)$ means the gaussian PDF centered on $\mu$ and width $\sigma$; $\u(a,b)$ is the uniform PDF between $a$ and $b$; $\logu(a,b)$ is the log-uniform PDF between $a$ and $b$. Dots indicate the parameters that were not directly sampled but were derived from the samples of the ``fitted" parameters.}
\end{deluxetable*}

\subsection{Possible Origins of the Difference with Previous Work}

The inference of a large stellar obliquity in \citet{2015ApJ...814...67A} was largely based on an apparently V-shaped transit light curve for KOI-89.02 (their Figure 4), which was argued not to be well fit by a high impact parameter transit. This lead them to obtain a solution involving a rapidly rotating, nearly pole-on star. Indeed, the pole-to-equator flux contrast induced by strong gravity darkening can make transits V-shaped, and a pole-on configuration was also required to explain the observed $v\sin i_\star$.
The solution also involved a moderate impact parameter so that the planet can travel across the regions on the stellar surface with significantly different surface brightness; this in turn led to a large eccentricity to explain the short observed transit duration. 

The crux of the disagreement is therefore that we do not find the stacked transit light curve of KOI-89.02 to be V-shaped (Figure \ref{fig:gdfit}, bottom). 
The origin of the difference of the derived transit shapes is unclear, but possible explanations include incorrect TTV corrections and distortion due to the binning of the light curve. 
Our stacked light curve, as well as individual transits of KOI-89.02, are well modeled by a high-impact parameter orbit and obviate the need for a large stellar obliquity.
In addition, we found that a large orbital eccentricity as required in the previous solution results in rapid orbital instability \citep[as was also pointed out in][]{2015ApJ...814...67A} and is not favored from a dynamical point of view (see Section \ref{sec:photod}).

\section{Summary and Discussion}\label{sec:discussion}

We performed a photodynamical analysis and transit shape modeling of the KOI-89/Kepler-462 system to constrain the masses and orbits of the two transiting planets (including the transiting KOI-89.02, which was previously classified as a candidate) and the spin of the host star. This is one of few multi-transiting systems confirmed around hot stars, for which Doppler masses are difficult to obtain. Our modeling of the transit light curves including mutual gravitational interactions indicate that the orbits of the two transiting planets have low eccentricities and a low mutual inclination. The modeling shows that the masses of both KOI-89.01/Kepler-462b and KOI-89.02 must be planetary, thus confirming KOI-89.02 to be a new planet (Kepler-462c). The large TTVs of KOI-89.02 suggest that either KOI-89.01/Kepler-462b is unusually massive given its size, or that there is a third planet that has evaded transit detection but is significantly perturbing the orbit of KOI-89.02. 

We also modeled the transit light curve of the two planets taking into account possible effects of gravity darkening of the rapidly rotating host star. We did not detect the signature of gravity darkening clearly, and concluded that stellar obliquity is not well constrained, as opposed to a previous claim. Below we discuss implications of those results.

\subsection{Large Stellar Obliquity: Nature or Nurture?}\label{ssec:discussion_model}

We focused on the KOI-89 system because it is one of three multi-transiting systems with claimed significant spin--orbit misalignments.
Such systems might imply that the inner regions of planetary systems are occasionally misaligned with the stellar spin, for reasons unrelated to dynamical evolution of planetary orbits (e.g., misaligned disk). 
As discussed in Section \ref{sec:intro}, the large stellar obliquity in the other two systems, Kepler-56 and HD~3167, could plausibly be explained by dynamical evolution of the planetary orbits after formation.
This left KOI-89 as the best known candidate for an obliquity excitation mechanism that does not alter planeatry orbits.

Our reanalysis of the KOI-89 data shows that there is no evidence for a high stellar obliquity as previously claimed by \citet{2015ApJ...814...67A}, although a high obliquity is not necessarily ruled out. 
We therefore conclude that there is currently no clear evidence for  spin--orbit misalignments unrelated to dynamical orbital evolution from the available obliquity constraints on stars with multiple transiting planets.

Obliquity measurements for a larger number of individual multi-transiting systems will serve as a benchmark for untangling the origin of high obliquities. 
This is particularly important for hot stars like KOI-89, given the statistical evidence that high obliquities may be the norm among stars hotter than $\approx 6000\,\mathrm{K}$, not only for those with hot Jupiters \citep{2010ApJ...719..602S, 2010ApJ...718L.145W, 2012ApJ...757...18A} but for other {\it Kepler} stars with smaller and/or longer-period planets \citep{2015ApJ...801....3M};
although \citet{2017AJ....154..270W} found otherwise and the reason of the apparent discrepancy remains unclear.
Given that the majority of the samples in these works are stars with single transiting planets, which may be a part of multi-planetary systems with misaligned orbits \citep{2018ApJ...860..101Z, 2019MNRAS.490.4575H}, these statistical results would be less sensitive to multi-transiting systems around hot stars. Obliquity constraints in multi-transiting systems would thus play a key role in uncovering the connection between the spin--orbit misalignment in generic {\it Kepler} systems around hot stars and dynamical excitation of planetary orbits.

\subsection{Masses and Radii of the KOI-89 Planets}\label{ssec:discussion_mass}

\begin{figure*}
\plotone{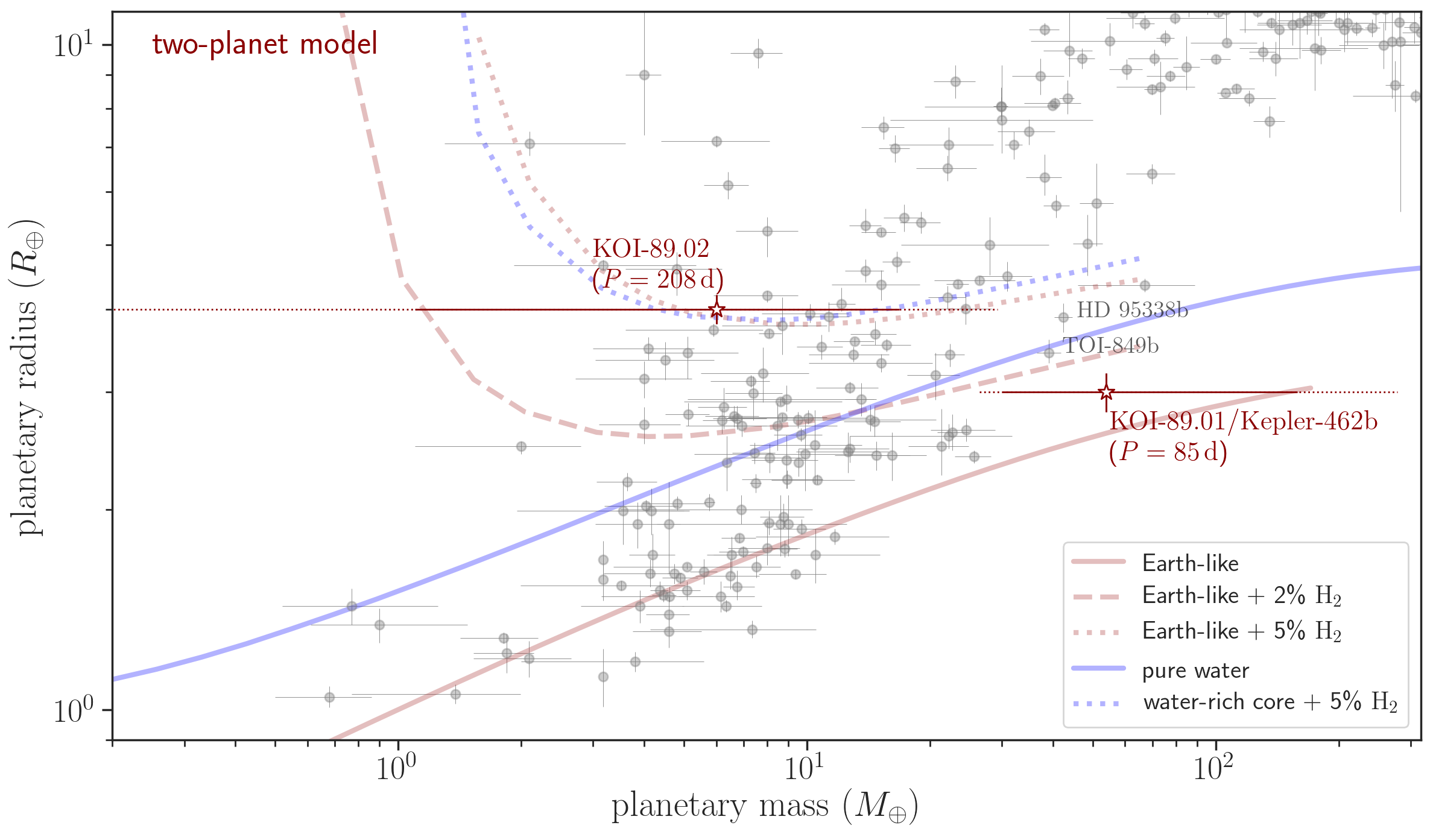}
\plotone{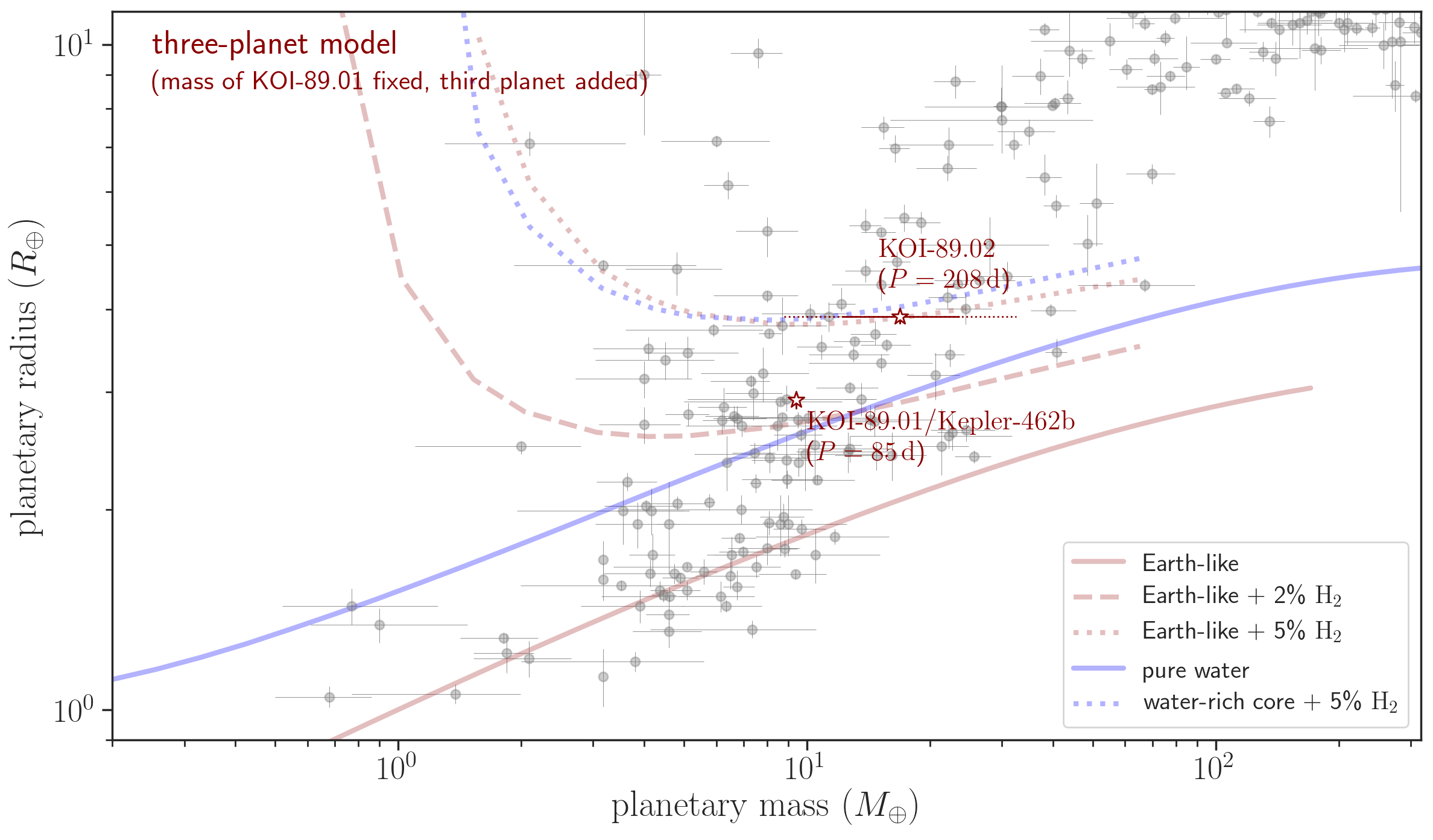}
\caption{Radius--mass diagram of planets smaller and less massive than Jupiter (gray circles with error bars). The selection criteria are similar to the ones in Figure \ref{fig:radius_period_smass}, except that  planets with masses from \citet{2014ApJ...787...80H} and \citet{2014ApJS..210...25X} are removed because those masses 
are derived using the analytic TTV formula \citep{2012ApJ...761..122L} without taking into account the dependence on free eccentricities.
 Star symbols with error bars show the estimated parameters of the KOI-89 planets, for the two-planet (top) and three-planet (bottom) photodynamical models (see Section \ref{sec:photod} for details). The locations of the symbols show the maximum a posteriori values, and solid (dotted) error bars correspond to 68\% (95\%) highest density intervals of the marginal posteriors. Solid, dashed, and dotted curves show theoretical radius--mass relations from \citet{2019PNAS..116.9723Z}.}
\label{fig:mr}
\end{figure*}

KOI-89.02 exhibits significant TTVs (Figure \ref{fig:ttvs}). The simplest explanation is that they are caused by KOI-89.01, and we showed that this model works well if KOI-89.01 is more massive than $\sim 20\,M_\oplus$. The value is unusually high given its size ($\approx 3\,R_\oplus$): such a planet is physically allowed, but we do not (yet) know of such planets (Figure \ref{fig:mr}, top). This motivated us to test the scenario in which the TTVs of KOI-89.02 are caused by another undetected planet. We found that this model also explains TTVs well for more ``natural" masses for the two transiting planets  (Figure \ref{fig:mr}, bottom) --- at the cost of increasing the model complexity to a level that is not justified by the data. It is in principle possible to quantify which model is favored from a Bayesian perspective, but the results would hardly be convincing given the lack of our current knowledge on the possible mass--radius relation, and on the plausible properties of the third planet conditioned on the properties of the star and two transiting planets. 

Nevertheless, we feel that the two-planet solution with a massive KOI-89.01 is worth some discussion in light of recent observations that found planets with similar properties (i.e., prior knowledge that assigns more plausibility to a model involving a dense sub-Neptune).
For example, \citet{2020Natur.583...39A} reported the discovery of TOI-849b from the {\it Transiting Exoplanet Survey Satellite (TESS)} data. It is an ultra-short-period (0.765524~days) planet with a mass of $39.1^{+2.7}_{-2.6}\,M_\oplus$ and a radius of $3.44^{+0.16}_{-0.12}\,R_\oplus$. The mass and radius indicate that the gaseous envelope, if any, should be very thin and \citet{2020Natur.583...39A} argued that KOI-849b may be the remnant core of a former giant planet. As shown in the top panel of Figure \ref{fig:mr}, the mass and radius of KOI-89.01 as inferred from the two-planet model is similar to TOI-849b, and imply that mass fraction of the gaseous envelope should be a few percent or less. If the two-planet model is correct, KOI-89.01, with an orbital period of $85\,\mathrm{days}$, would suggest that such a massive core without a thick envelope can form further away from the star where the effect of photoevaporation is less significant. This poses a theoretical challenge similar to the one posed by super Earths: why would such a planet not grow into a gas giant? This property is even more puzzling in the presence of the outer less massive KOI-89.02 with a thicker atmosphere (as simply implied by its larger radius), whose core would have had a longer formation timescale than the inner KOI-89.01, all else being equal. 
Such a large contrast in the mean densities has been observed in the Kepler-36 system \citep{2012Sci...337..556C} and has been explained by photoevaporation \citep[e.g.,][]{2013ApJ...776....2L}, but the planets of KOI-89 are much further away from the star where this explanation would not work.
This might then suggest that accretion onto planetary cores significantly depends on the surrounding environment. 

Another implication of the two-planet solution is that such dense, (sub-)Neptune-sized planets may be abundant on orbits wider than have been probed with Doppler surveys to date. If so, Doppler follow-up observations of relatively long-period transiting Neptune-sized planets orbiting bright stars from {\it K2} or {\it TESS} would reveal more such planets. HD 95338 b \citep{2020MNRAS.496.4330D}, a planet with $42.4^{+2.2}_{-2.1}\,M_\oplus$ and $3.89^{+0.19}_{-0.20}\,R_\oplus$ (Figure \ref{fig:mr}, top) on a 55-day period orbit around a bright K dwarf observed by {\it TESS}, may be such an example.

\acknowledgments 

The authors thank the referee, Jason Barnes, for his constructive comments which helped to improve the clarity of this paper. KM gratefully acknowledges the support by the W.\,M.~Keck Foundation Fund at the Institute for Advanced Study, where a part of this work was performed. Support for DT was provided by NASA through the NASA Hubble Fellowship grant HST-HF2-51423.001-A awarded  by  the  Space  Telescope  Science  Institute,  which  is  operated  by  the  Association  of  Universities  for  Research  in  Astronomy,  Inc.,  for  NASA,  under  contract  NAS5-26555.
Simulations in this paper made use of the {\tt REBOUND} code which is freely available at \url{http://github.com/hannorein/rebound}.

\appendix

\section{Light-curve Analysis}

\subsection{Modeling of the Stacked Transit Light Curves with a Normal Transit Model}\label{ssec:phase}

Here we present modeling of the stacked transit light curves as processed in Section \ref{ssec:photod_data} using the model for a quadratically limb-darkened star \citep{exoplanet:luger18}. We do not include the effect of gravity darkening, and fit for orbital eccentricities without incorporating the prior knowledge from photodynamical modeling. The analysis here is meant to illustrate the possible range of solutions prior to the modeling of dynamical interactions.

\begin{figure*}
	\epsscale{1.15}
	\plottwo{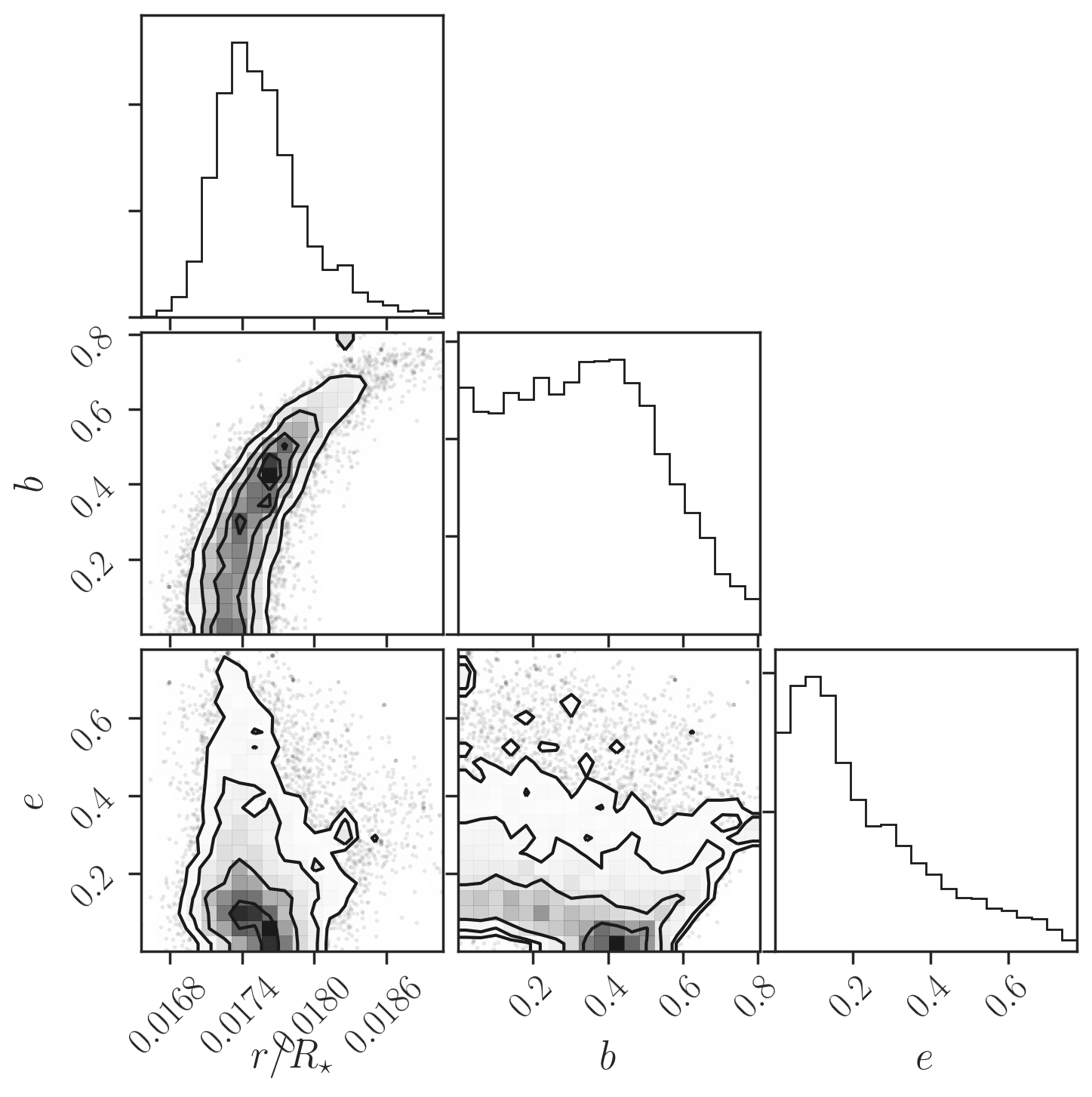}{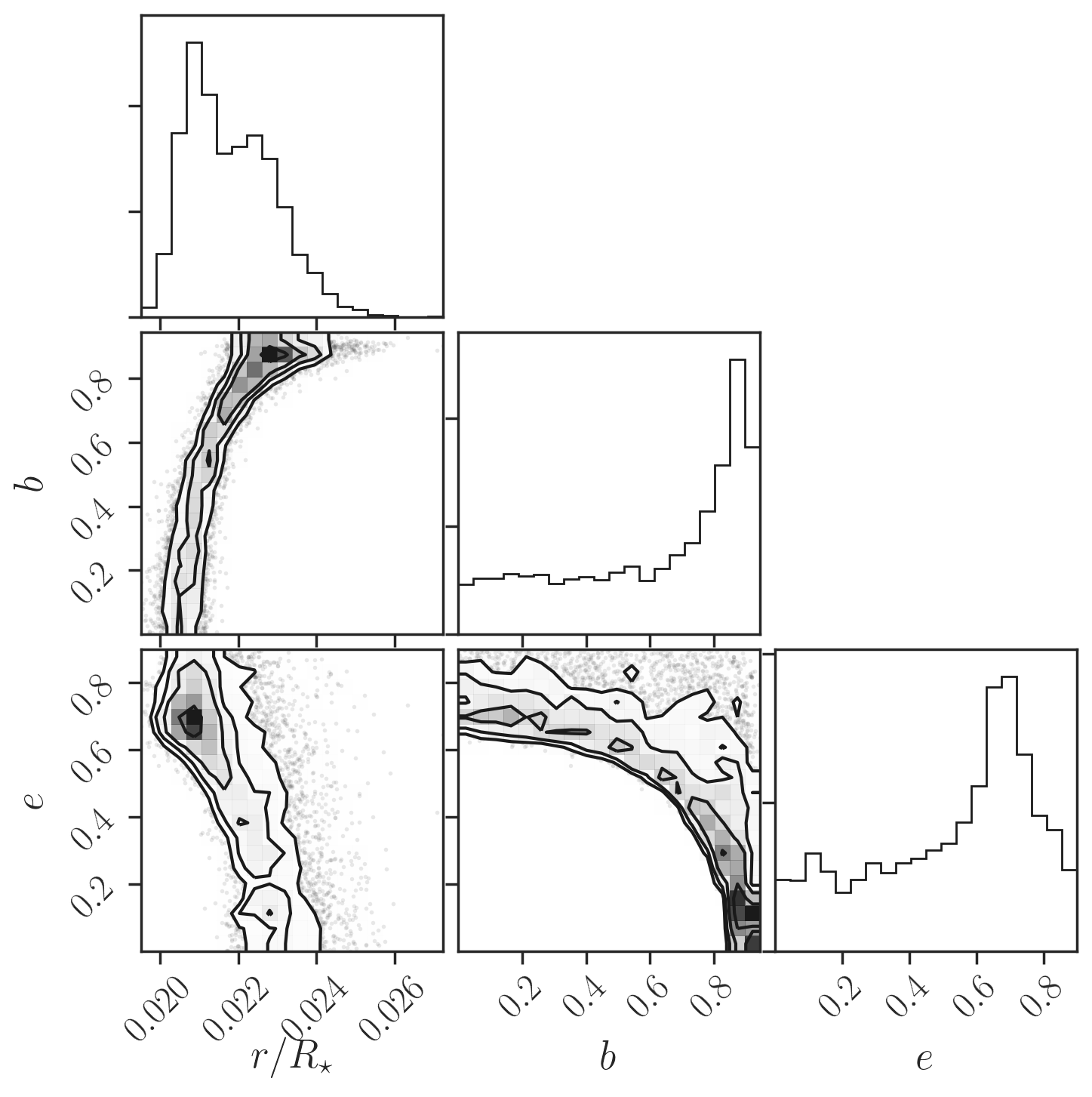}
	\caption{Corner plots for the posterior samples of radius ratios, impact parameters, and eccentricities of KOI-89.01 (left) KOI-89.02 (right).}
	\label{fig:phasefit}
\end{figure*}

The modeling was performed using \textsf{exoplanet} \citep{exoplanet} and its dependencies \citep{exoplanet:agol19, exoplanet:astropy13, exoplanet:astropy18, 2013MNRAS.435.2152K, exoplanet:luger18, exoplanet:pymc3,exoplanet:theano}. The posterior samples were obtained for the parameters listed in Table \ref{tab:koi89_phase} using the prior in the rightmost column. The noise was modeled using the Gaussian process kernel in Equation \ref{eq:kernel}, and the light curve of each planet was modeled individually. Table \ref{tab:koi89_phase} summarizes the resulting HDIs for marginal posteriors, and Figure \ref{fig:phasefit} shows corner plots \citep{corner} for the radius ratio, impact parameter, and eccentricity. 

For the radius ratios, we find good agreement with the results of photodynamical modeling in Section \ref{sec:photod} using unstacked light curves, and gravity-darkened modeling of stacked light curves in Section \ref{sec:gravity_darkening} that adopted a gravity-darkened model. On the other hand, eccentricity and impact parameter are poorly constrained based only on the shapes of the stacked light curves and the prior on the mean stellar density, and these parameters exhibit strong correlations as shown in Figure \ref{fig:phasefit}. We note that the marginal posterior for the eccentricity of KOI-89.02 is peaked at a high value, but lower values are also allowed. The former is the solutions found by \citet{2015ApJ...814...67A}, and we found that the latter should be the case from photodynamical modeling in Section \ref{sec:photod}.
Such a posterior is found because the duration of KOI-89.02 is relatively short for the given orbital period and prior on the mean density ($\sim$ stellar radius), and because the durations of ingress/egress are not well constrained by the data: here the short duration can be explained either by a large, finely tuned $b$ and a low $e$, or a high $e$ and a wider range of $b$. The latter solution has a larger volume in the parameter space and produces a peak in the marginal posterior for $e$. This example illustrates the importance of carefully interpreting the short-duration transits. 

\subsection{Central Times and Durations of Individual Transits}\label{ssec:ttv}

We also fitted individual transits separately using {\tt exoplanet} with the same noise model. This time we fixed $e=0$ and set log-uniform prior on the mean stellar density, because we are interested only in empirically constraining mid-transit times and transit durations calculated as $T=(R_\star P/\pi a)\sqrt{1-b^2}$ \citep{2010arXiv1001.2010W}, so that TTVs and TDVs can be visualized and compared to the results of the full photodynamical modeling.
Here we also modeled the overlapping transit of the two planets around $\mathrm{BJD}_\mathrm{TDB}=2454833+912.7$, so that the values can be used to test predictions based on the other transits used for the photodynamical modeling. The resulting transit times and durations (medians and symmetric $68\%$ intervals of the marginal posteriors) are shown in Figure \ref{fig:ttvs}. We emphasize again that these transit times and durations were not used for any modeling, but are compared with the output of the photodynamical modeling in Section \ref{sec:photod} only for an illustrative purpose. 

\begin{deluxetable*}{l@{\hspace{.5cm}}cc@{\hspace{.5cm}}cc@{\hspace{.5cm}}c}[!ht]
\tablecaption{Parameters from Phase-folded Transit Light Curves.\label{tab:koi89_phase}}
\tablehead{
\colhead{} & \multicolumn{2}{c}{KOI-89.01} & \multicolumn{2}{c}{KOI-89.02}\\
\colhead{} & \colhead{68\% HDI} & \colhead{$95\%$ HDI} & \colhead{68\% HDI} & \colhead{$95\%$ HDI} & \colhead{prior}}
\startdata
time of inferior conjunction $t_0$ (days) &  $[-0.001, 0.001]$ &  $[-0.002, 0.003]$  &  $[-0.001, 0.002]$ &  $[-0.002, 0.003]$   & $\g(0,0.05)$\\
radius ratio $r/R_\star$ &  $[0.0171, 0.0178]$ &  $[0.0169, 0.0183]$  &  $[0.0205, 0.0227]$ &  $[0.0201, 0.0240]$ & $\logu(0.001, 0.05)$\\
impact parameter $b$ &  $[0.09, 0.51]$ &  $[0.00, 0.66]$  &  $[0.46, 0.92]$ &  $[0.07, 0.92]$   & $\u(0, 1+r/R_\star)$\\
eccentricity $e$ &  $[0.00, 0.28]$ &  $[0.00, 0.61]$  &  $[0.40, 0.86]$ &  $[0.07, 0.89]$ & $\u(0,0.9)$\\
argument of periastron $\omega$ (deg) &  $[-35, 180]$ &  $[-180, 169]$  &  $[8, 151]$ &  $[-158, 180]$   & $\u(-180,180)$ \\
mean stellar density $\rho_\star$ ($\mathrm{g\,cm^{-3}}$) &  $[0.47, 0.59]$ &  $[0.42, 0.65]$  &  $[0.47, 0.59]$ &  $[0.43, 0.66]$  & $\g(0.53, 0.06)$ \\
limb-darkening coefficients $(u_1+u_2)^2$ &  $[0.23, 0.54]$ &  $[0.13, 0.80]$  &  $[0.06, 0.30]$ &  $[0.00, 0.59]$   & $\u(0,1)$\\
limb-darkening coefficients $u_1/2(u_1+u_2)$ &  $[0.00, 0.20]$ &  $[0.00, 0.48]$  &  $[0.00, 0.44]$ &  $[0.00, 0.87]$   & $\u(0,1)$\\
mean flux ($10^{-5}$) &  $[-0.3, 0.4]$ &  $[-1.1, 1.2]$  &  $[-0.5, 0.6]$ &  $[-1.5, 1.5]$ & $\g(0,10^3)$ \\
$\ln \sigma_{\rm jit}$ &  $[-10.5, -10.3]$ &  $[-10.6, -10.3]$  &  $[-12.4, -10.9]$ &  $[-13.0, -10.7]$  & $\u(-13, -7)$ \\
$\ln\alpha$ &  $[-12.6, -11.4]$ &  $[-13.0, -10.9]$  &  $[-12.0, -10.7]$ &  $[-13.0, -10.6]$ & $\u(-13, -7)$   \\
$\ln \rho$ (days) &  $[-4.1, -0.8]$ &  $[-5.0, 0.4]$  &  $[-5.0, -2.1]$ &  $[-5.0, 0.4]$  & $\u(-5,1)$ \\
\enddata
\tablecomments{Values listed here are the 68\%/95\% highest density intervals (HDIs) of the marginal posteriors. Priors --- $\mathcal{N}(\mu,\sigma)$ means the gaussian PDF centered on $\mu$ and width $\sigma$; $\u(a,b)$ is the uniform PDF between $a$ and $b$; $\logu(a,b)$ is the log-uniform PDF between $a$ and $b$.}
\end{deluxetable*}

\section{Constraints on Mutual Orbital Inclinations in the Two-planet Model with Low Eccentricities}\label{sec:misaligned}

\begin{figure}
	\epsscale{1.15}
	\plotone{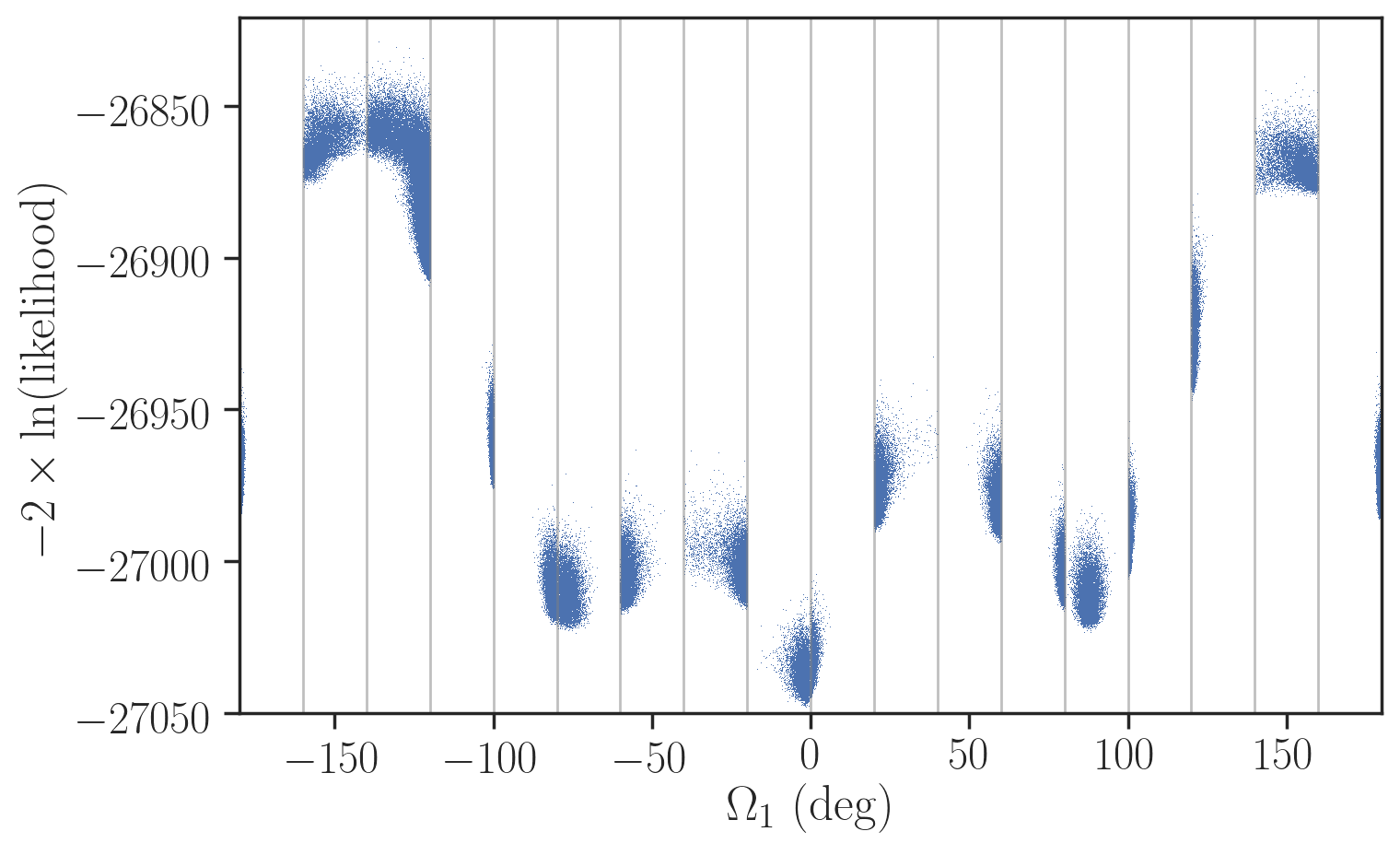}
	\caption{Dependence of the likelihood on $\Omega_1$ in the two-planet model with low-eccentricity orbits.}
	\label{fig:loglike_domega}
\end{figure}

\begin{figure}
	\epsscale{1.15}
	\plotone{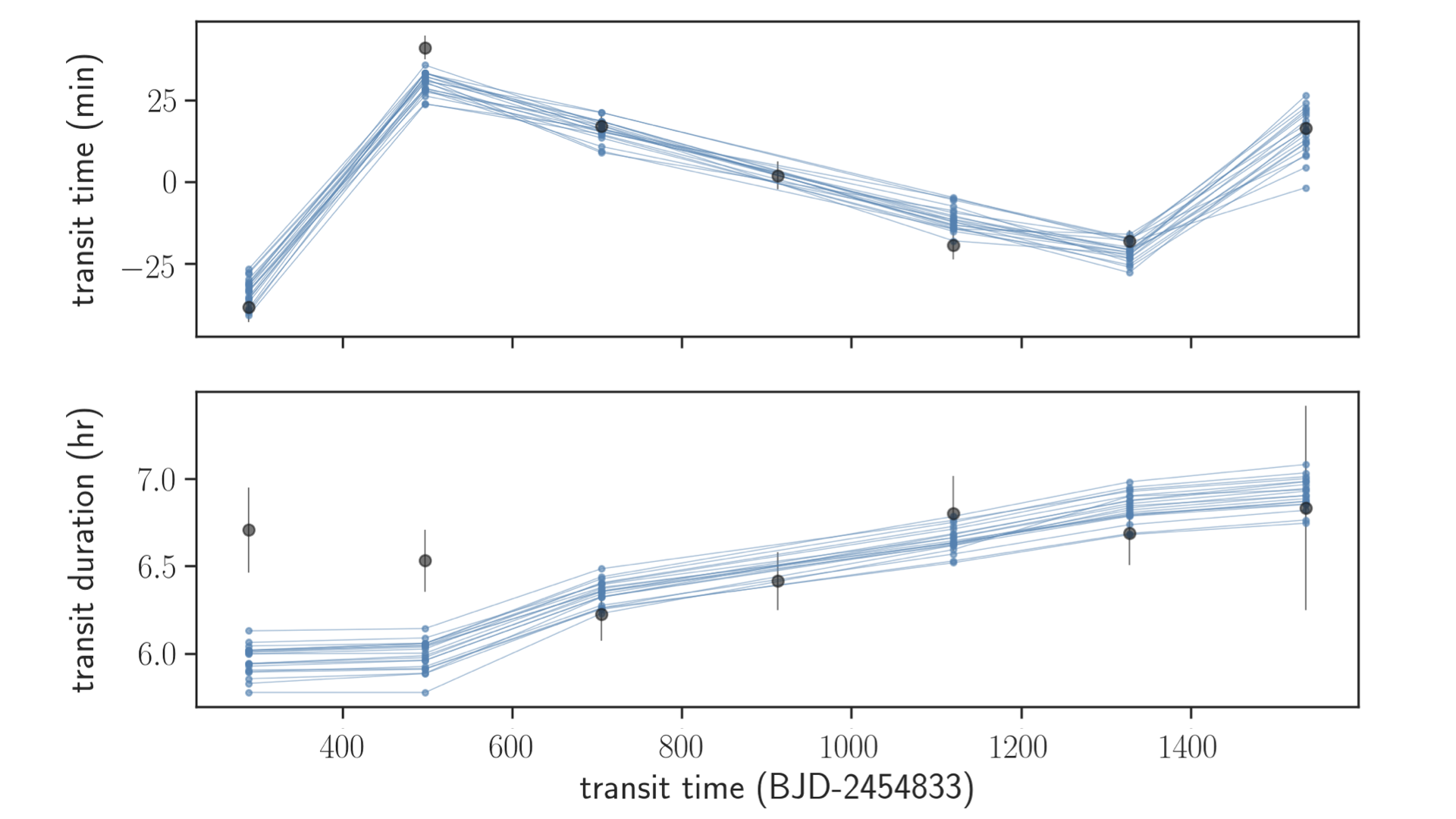}
	\plotone{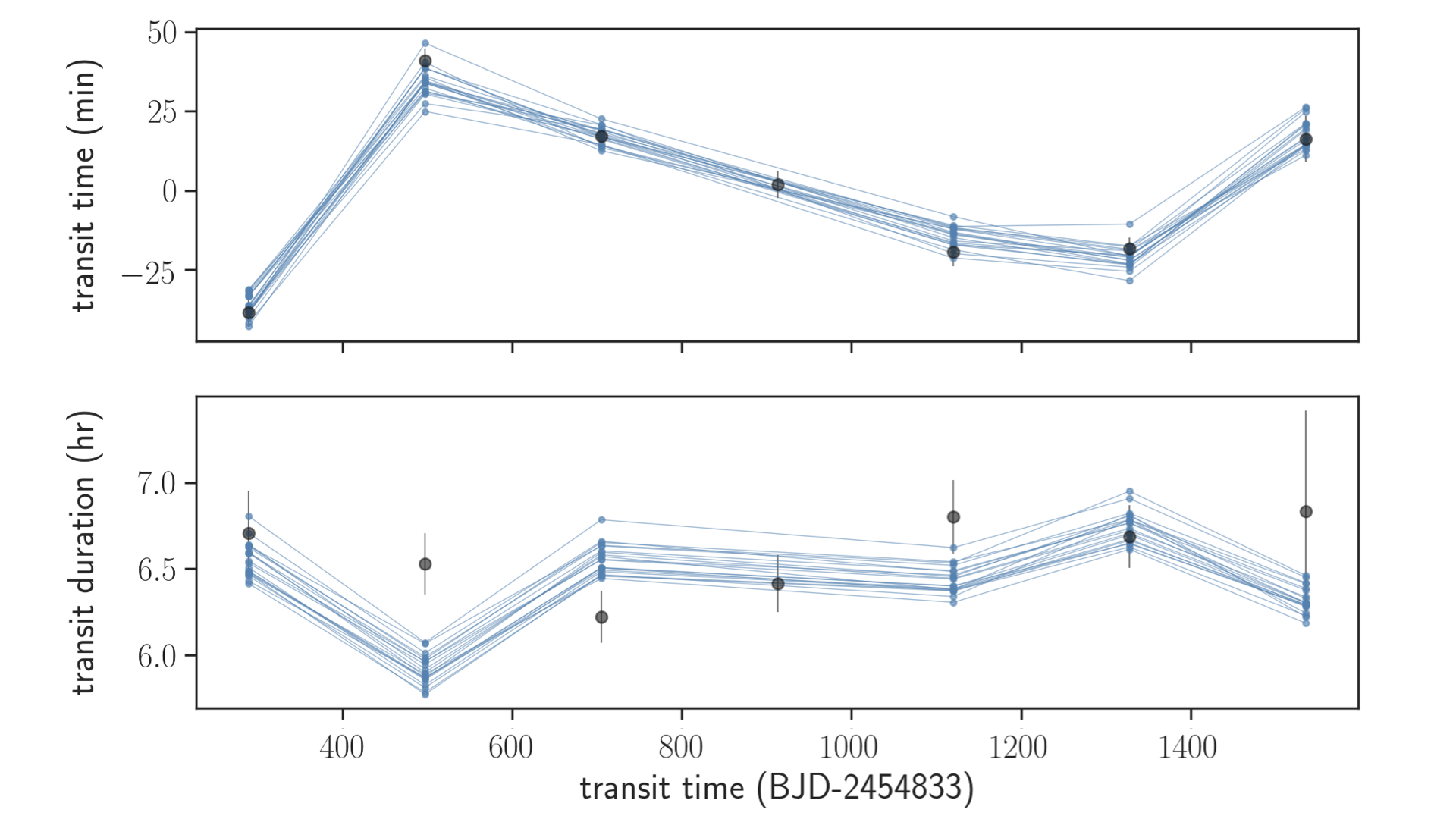}
	\caption{TTV and TDV solutions of KOI-89.02 for the fits restricted to $\Omega_1=[-40^\circ,-20^\circ]$ (top) and $\Omega_1=[-100^\circ,-80^\circ]$ (bottom).}
	\label{fig:grid}
\end{figure}

In Section \ref{ssec:photod_two}, we argued that two-body models with low-eccentricity, highly inclined orbits are disfavored by the lack of TDVs. Here we show how the goodness of fit depends on the mutual orbital inclination, which in our case is essentially the same as the orbital misalignment in the sky plane, $\Omega_1$. We repeated the posterior sampling with uniform priors on the mass ratios as in Section \ref{ssec:photod_two}, separately for 18 equally-spaced intervals of $\Omega_1$. Figure \ref{fig:loglike_domega} shows the resulting posterior samples with the log-likelihood values (multiplied by $-2$ so that they analogize with chi-squared), where the vertical lines indicate the boundaries of the $\Omega_1$ intervals. As we argued in the main text, the best solution is around $\Omega_1=0^\circ$. Those solutions with intermediate $|\Omega_1|\sim30^\circ$ give a bad fit because the durations of KOI-89.02 drift too much (Figure \ref{fig:grid}, top). There exist local minima around $|\Omega_1|\sim90^\circ$ because the duration variations due to nodal precession are minimized for nearly orthogonal orbits. However, in this case KOI-89.02 receives kicks perpendicular to its orbit and its durations still fluctuate more than allowed by the data (Figure \ref{fig:grid}, bottom).
We note that better solutions than shown in Figure \ref{fig:loglike_domega} exist for most values of $\Omega_1\neq0$ if larger values of eccentricity are considered for KOI-89.02. However, those solutions are disfavored from the stability point of view, as discussed in Section \ref{sssec:photod_two_eccentric}.

\bibliographystyle{aasjournal}



\listofchanges

\end{document}